\documentclass[aoas,preprint]{imsart}

\usepackage{amsthm,amsmath,natbib}
\usepackage[colorlinks,citecolor=blue,urlcolor=blue]{hyperref}
\usepackage{amsfonts,amssymb}

\usepackage{graphicx}
\usepackage{wrapfig}
\usepackage[font={small,it}]{caption}
\usepackage{array}
\usepackage{soul}
\arxiv{arXiv:1506.05244}

\startlocaldefs
\usepackage{mathtools}
\usepackage{relsize}

\endlocaldefs

\begin{document}

\begin{frontmatter}

% "Title of the paper"
\title{Detection of Epigenomic Network Community Oncomarkers}
\runtitle{Epigenomic Network Community Oncomarkers}

% indicate corresponding author with \corref{}
\begin{aug}
\author{\fnms{Thomas E.} \snm{Bartlett}\thanksref{t1,t2,m1}\ead[label=e1]{thomas.bartlett.10@ucl.ac.uk}},
\and
\author{\fnms{Alexey} \snm{Zaikin}\thanksref{t3,m1}}

\thankstext{t1}{thomas.bartlett.10@ucl.ac.uk}
\thankstext{t2}{TB acknowledges funding from EPSRC grant no. EP/M507970/1 and previous support from EPSRC and MRC via UCL CoMPLEX.}
\thankstext{t3}{AZ acknowledges support from the Deanship of Scientifc Research (DSR), King Abdulaziz University (KAU), Jeddah, under grant No. (86-130-35-RG) and from the Russian Foundation for Basic Research (14-02-01202,13-02-00918).}
\runauthor{T.E. Bartlett et al}

\affiliation{University College London\thanksmark{m1}}

\end{aug}

\begin{abstract}
In this paper we propose network methodology to infer prognostic cancer biomarkers based on the epigenetic pattern DNA methylation. Epigenetic processes such as DNA methylation reflect environmental risk factors, and are increasingly recognised for their fundamental role in diseases such as cancer. DNA methylation is a gene-regulatory pattern, and hence provides a means by which to assess genomic regulatory interactions. Network models are a natural way to represent and analyse groups of such interactions. The utility of network models also increases as the quantity of data and number of variables increase, making them increasingly relevant to large-scale genomic studies. We propose methodology to infer prognostic genomic networks from a DNA methylation-based measure of genomic interaction and association. We then show how to identify prognostic biomarkers from such networks, which we term `network community oncomarkers'. We illustrate the power of our proposed methodology in the context of a large publicly available breast cancer dataset.
\end{abstract}

\begin{keyword}
\kwd{Computational biology}
\kwd{stochastic networks}
\kwd{community detection}
\kwd{epigenomics.}
\end{keyword}

\end{frontmatter}

\section{Introduction}
Complex systems which can be modelled as networks are ubiquitous. Well-known examples include social/communication networks \citep{beguerisse2014interest} and economic networks \citep{saavedra2014structurally}, as well as many others in the biological sciences such as ecological networks \citep{nandi2014social}, gene networks \citep{wei2010network,li2014quantifying}, protein networks \citep{mardia2013statistical,tran2013relationship}, and metabolic networks \citep{reznik2013stubborn}. Over the past few years in cell biology, much focus has shifted from investigation of individual genes, to pathways of genes, to gene networks. The interest in novel methodology for network analysis in cell biology follows from the recognition that examining the way genes work in groups often yields more accurate inference of biological processes.

The problem of finding community structure in networks has been studied for many years. Important applications of this problem include identifying groups of friends or co-workers in social networks, as well as identifying functional subnetwork modules in biological networks \citep{girvan2002community}. In the biological setting, genes can be viewed as acting together as part of `subnetwork modules', which are functional units with specific biological roles \citep{shen2002network}. Indeed, it has been demonstrated recently that such modularity is a natural and even inevitable result of evolutionary pressures \citep{clune2013evolutionary}. This is because modularity minimises network connectivity cost whilst maximising performance, and thus it represents the most parsimonious and efficient type of network structure for biological networks such as these. Furthermore, considering groups of genes defined together as subgraphs can lead to big increases in statistical power, aiding discovery of biological phenomena \citep{jacob2012more,li2010variable,peng2010regularized}. Therefore, it is relevant to both the biological and statistical modelling to consider the group behaviour of genes in this way. Hence, this viewpoint of modular genomic network structure is fundamental to the methodology we propose here.

Epigenetic patterns are gene-regulatory patterns, meaning that they influence the activity of particular genes, among other phenomena \citep{jones2012functions}. Epigenetic information can be modulated during the lifetime of an organism by environmental cues \citep{feinberg2006epigenetic,cooney2007epigenetics,christensen2009epigenetic}. As such, epigenetics can be considered to be an interface between the genome and the environment, and consequently also a conduit for environmental risk factors. Alterations in the epigenetic pattern DNA methylation are among the earliest changes in human carcinogenesis \citep{feinberg2006epigenetic}, and hence DNA methylation patterns are expected to yield important prognostic information useful for biomarker development. DNA methylation patterns are thought promising for biomarker development in a wide variety of physiological systems and organs \citep{verschuur2012epigenetic,van2013assessment,fleischer2014genome,kishida2012epigenetic,gao2013dna,kang2001cpg,kang2003profile,bhagat2012aberrant,yamamoto2012molecular,luo2014differences,navarro2012genome,maekawa2013genome}. 

It is well established that DNA methylation plays an important role in gene regulation, and hence DNA methylation patterns often reflect gene regulatory behaviour \citep{jones2012functions}. Changes in DNA methylation are highly stochastic. The timescale over which these changes take place is much faster than DNA mutations can arise, but much slower than the transient and periodically varying activity of individual genes, and this timescale is ideal for biomarker development. DNA methylation data are extremely noisy; however, statistics which summarise DNA methylation patterns at the gene level have been shown to have much utility as analytical tools \citep{bartlett2013corruption}. It has been shown previously that DNA methylation can serve as a surrogate measure of genomic-regulatory action \citep{brocks2014intratumor}. Hence, DNA methylation measurements are a natural basis from which to construct genomic regulatory and related networks. As a cancer progresses, its signalling and control networks are rearranged (`rewired'), leading to genomic changes which are advantageous for the cancer \citep{barabasi2004network}. Previous research has found that patient survival outcome in breast cancer can be predicted well by network models of this rewiring, based on gene expression data \citep{taylor2009dynamic}. Hence, network models based on DNA methylation measurements are a very promising basis for the development of prognostic biomarkers.

Statistical network models are a parsimonious way to represent and analyse large numbers of variables and samples. They are efficient analytical tools appropriate for the very large datasets which are produced by the latest technologies in cell biology. When carrying out modelling of this type, it is important to balance statistical fidelity with computational efficiency. The `stochastic blockmodel' (SBM) \citep{holland1983stochastic,bickel2009nonparametric} is an efficient network model which has been widely studied and is well understood, and hence it is a good basis for our proposed methodology. Under the SBM, there is a greater probability of observing an edge (or interaction) between a pair of nodes if they are in the same block, or community. The Newman-Girvan modularity \citep{newman2004finding} quantifies the extent to which network edges are observed between community members, for a particular assignment of nodes to communities, compared to the expected number of edges between community members if there were no community structure present. It can be shown that, under certain conditions, fitting the stochastic blockmodel is equivalent to maximising the Newman-Girvan modularity over a network, and that these are both equivalent to spectral clustering \citep{riolo2012first,bickel2009nonparametric}. We use spectral clustering as an efficient computational algorithm for fitting the SBM.

It has also been shown recently that, under reasonable assumptions, the SBM can be used to represent any network as a `network histogram', whatever the generating mechanism of that network. Further, the network histogram provides a heuristic method to estimate the optimum number of blocks, or clusters, which a valid blockmodel representation of the network may contain. This is important and useful, because it means that the blockmodel can be used to identify an unknown number of communities, or functional subnetwork modules, in a biological network. Genomic networks are typically scale-free, which means that they exhibit a power-law degree distribution \citep{wagner2002estimating}. Further, they are thought to be hierarchical \citep{barabasi2004network,palla2010multifractal}, displaying multi-scale properties. This means that different functional organisation is visible at different granularities, or scales. We use the network histogram method \citep{olhede2014net} to estimate the optimal granularity at which to identify communities, or functional subnetwork modules, in our prognostic networks by fitting the SBM.

The main contribution of this work is to propose a well-integrated, and well-validated, statistical methodology for detecting biomarkers from the biological viewpoint of modular genomic network structure, using DNA-based measurements of genomic regulatory patterns. To do this, we show how to integrate our previously proposed DNA methylation-based measure of interaction or association between pairs of genes, the `DNA methylation network interaction measure' \citep{bartlett2014dna}, into a multi-stage pipeline to construct prognostic network community-based biomarkers. This leads to our novel and generally applicable statistical methodology; we present the multiple stages of this methodology sequentially here, and thus this paper is organised as follows. In Section \ref{methodsSect}, we outline our previously proposed DNA methylation network interaction measure \citep{bartlett2014dna}, and we show how to use this measure to infer prognostic genomic networks. An edge between a pair of genes/nodes in these networks indicates that the strength of interaction or association between those genes is associated with disease progression. Also, in Section \ref{methodsSect}, we show how to identify prognostic biomarkers from such networks, using community detection to identify subnetwork modules within the network. These communities are groups of nodes/genes among which there is a high density of prognostic interactive or associative behaviour, and we term them `network community oncomarkers'. In Section \ref{resultsSect}, we demonstrate the utility of our proposed methodology in the context of a large, publicly available breast cancer dataset. To do so, we use each network community oncomarker to calculate a one-number prognostic score for each patient, and we use these scores to classify patients one by one into prognostic groups. Also in Section \ref{resultsSect}, we show that among the genes of the network community oncomarkers, the DNA methylation network interaction measure is associated with co-regulatory behaviour as measured by gene expression, justifying these findings in terms of biological function.

\section{Proposed methodology}\label{methodsSect}
\begin{figure}[!ht]
\centering\includegraphics[width=0.8\textwidth]{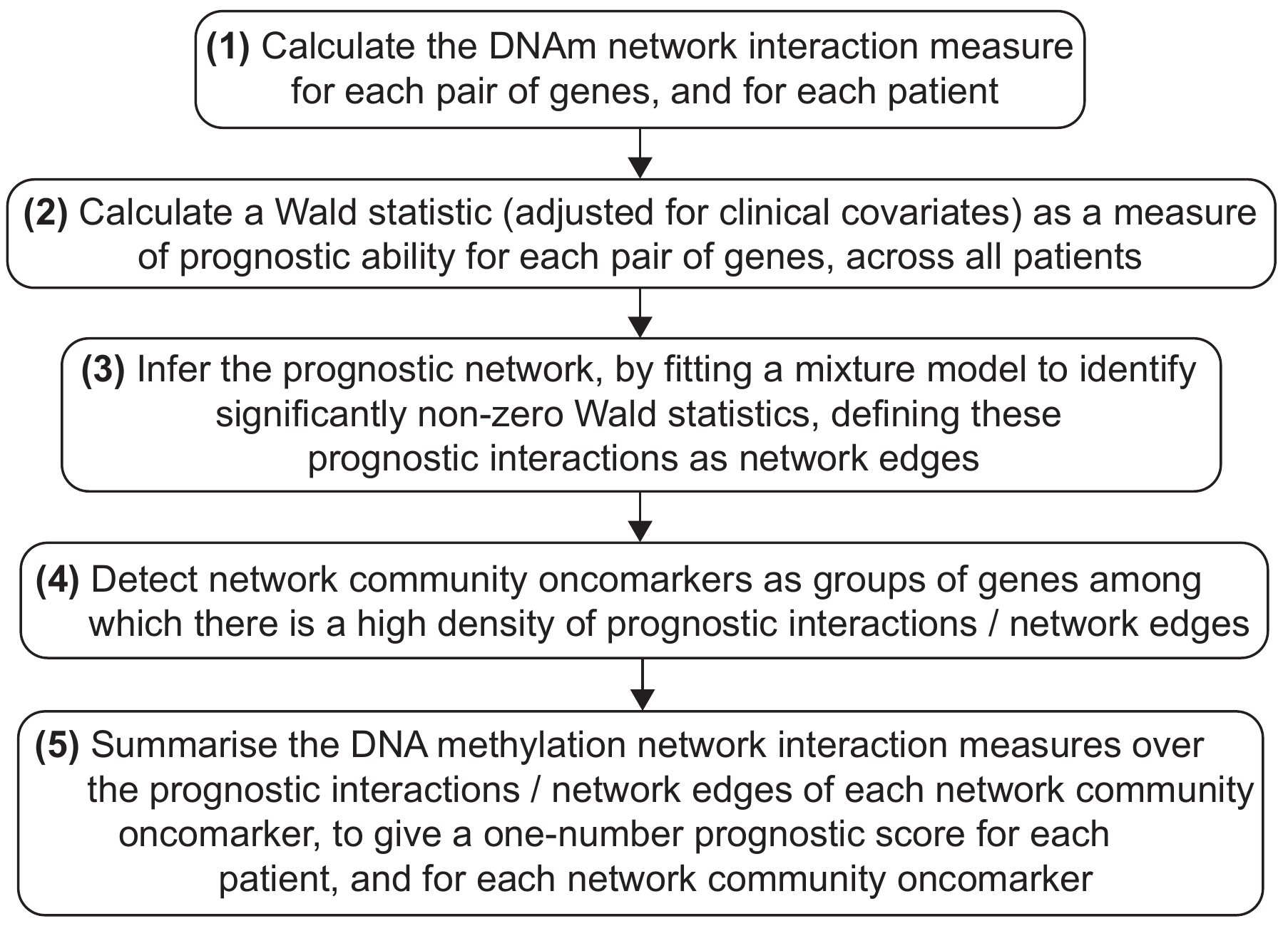}
\caption{Overview of methods.}\label{methodsFlowchart}
\vspace{-2ex}
\end{figure}

An overview of our proposed methodology appears in Figure \ref{methodsFlowchart}, following which component parts of this methodology are described in detail.

\noindent We note that, in principle, each of the steps illustrated in Figure \ref{methodsFlowchart} could be replaced with alternative choices of methodology.

\subsection{DNA Methylation Network Interaction Measure}\label{DNAmRhoSect}
\begin{figure}
\centering\includegraphics[width=1\textwidth]{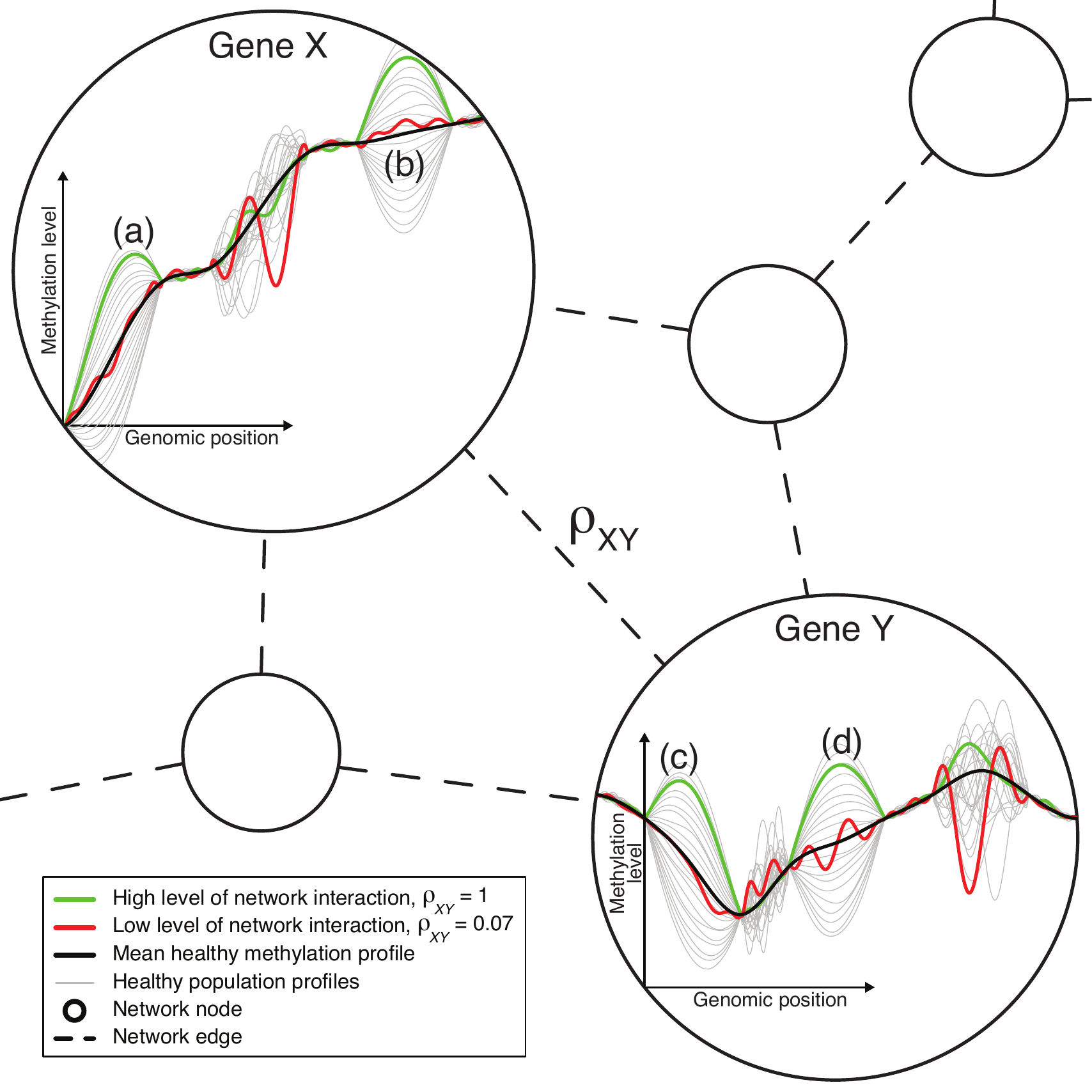}
\caption{The DNA methylation network interaction measure.}\label{DNAmMeasure}
\vspace{-3ex}
\caption*{A combination of the variation of the healthy methylation profiles in regions (a) and (b) of gene X explains well / is well-explained by a combination of the variation of the healthy methylation profiles in regions (c) and (d) of gene Y. The green cancer sample varies by a large amount about the mean methylation profile and in a typical way in these regions in both genes. Hence, the green sample corresponds to a high level of network interaction for this sample, $\rho_{XY}=1$. The equivalent variations in the other regions of these genes do not explain each other well, and so the red sample, which varies by a large amount in these other regions and varies less and in an atypical way in regions (a) - (d), corresponds to a low level of network interaction, $\rho_{XY}=0.07$. Genes X and Y are likely to have different numbers of methylation measurement locations (i.e., variables X and Y are of different dimension). The ordering of the measurement locations has no influence on the calculation of $\rho$, as long as the ordering is consistent across samples. This diagram was presented previously by \cite{bartlett2014dna}.}
\end{figure}

DNA methylation is a chemical modification to DNA which may occur at numerous locations within a gene: the pattern of these modifications within a gene forms a `DNA methylation profile'. Using canonical correlation analysis (CCA) \citep{hotelling1936relations} we previously proposed a statistic \citep{bartlett2014dna} which measures the strength of interaction or association between a pair of genes (network nodes) in a single sample/patient, based on DNA methylation profiles (Figure \ref{DNAmMeasure}). This statistic quantifies the extent to which the DNA methylation profiles of a pair of genes explain each other. It is based only on measurements of the DNA methylation profiles of that pair of genes, and it acts as a surrogate for a measure of the extent to which this pair of genes behave interactively or associatively. Such behaviour may include transcriptional regulation or co-regulation, or other types of biochemical interaction, influencing gene expression levels, isoforms and the presence of alternatively spliced gene products, among other phenomena \citep{jones2012functions}. The details of this DNA methylation network interaction measure are as follows.

The DNA methylation network interaction measure is defined by analogy to CCA. CCA aims to discover linear combinations of variables of one type, and linear combinations of variables of another type, so that these combinations best explain each other. In this context, a particular way of combining (by scaling and adding) the deviations from the mean methylation profile at a number of locations within one gene might be particularly effective at explaining a particular combination of (again, by scaling and adding) the deviations from the mean methylation profile at a number of locations in another gene, and vice versa. There will probably be fewer ways in which the methylation levels of these genes covary across the samples than there are locations at which methylation is measured along the genes; this is because the methylation level is highly correlated at many locations along a particular gene. CCA finds the most important components of this covariation across samples.

CCA seeks to find the vectors $a$ and $b$, in the $p$ and $q$ dimensional spaces of variables $\mathbf{X}=(x_1,x_2,...,x_p)'$ and $\mathbf{Y}=(y_1,y_2,...,y_q)'$, respectively, which maximise the correlation $\rho=\text{cor}\left(\mathbf{a}'\mathbf{X},\mathbf{b}'\mathbf{Y}\right)$ defined according to equation \ref{CCA}:
\begin{equation}
\rho=\frac{\mathbf{a}'\boldsymbol{\Sigma}_{XY}\mathbf{b}}{\sqrt{\mathbf{a}'\boldsymbol{\Sigma}_{XX}\mathbf{a}}\sqrt{\mathbf{b}'\boldsymbol{\Sigma}_{YY}\mathbf{b}}},\label{CCA}
\end{equation}
where
\begin{equation*}
\boldsymbol{\Sigma}_{XX}=\mathbb{E}\left[(\mathbf{X}-\boldsymbol{\mu}_X)(\mathbf{X}-\boldsymbol{\mu}_X)'\right]
\end{equation*}
and
\begin{equation*}
\boldsymbol{\Sigma}_{YY}=\mathbb{E}\left[(\mathbf{Y}-\boldsymbol{\mu}_Y)(\mathbf{Y}-\boldsymbol{\mu}_Y)'\right]
\end{equation*}
are the covariance matrices of $\mathbf{X}$ and $\mathbf{Y}$ respectively,
\begin{equation*}
\boldsymbol{\Sigma}_{XY}=\mathbb{E}\left[(\mathbf{X}-\boldsymbol{\mu}_X)(\mathbf{Y}-\boldsymbol{\mu}_Y)'\right]
\end{equation*}
is the cross-covariance matrix of $\mathbf{X}$ and $\mathbf{Y}$, and $\boldsymbol{\mu}_X$ and $\boldsymbol{\mu}_Y$ are the mean vectors of $\mathbf{X}$ and $\mathbf{Y}$.

Two genes $X$ and $Y$ have corresponding methylation profiles which are measured for sample / patient $k$ at $p$ and $q$ CpGs (loci) respectively along these genes. Denoting these measurements by the variables $x_1,...x_p$ and $y_1,...,y_q$ for genes $X$ and $Y$ respectively, the DNA methylation profiles for these genes, for patient $k$, can be represented by the vectors $\mathbf{x}(k)$ and $\mathbf{y}(k)$, which have $p$ and $q$ entries respectively. A measure of DNA methylation network interaction $\rho_{XY}(k)$, of the methylation profiles of genes $X$ and $Y$ for sample $k$, can then be defined by analogy with equation \ref{CCA}, according to equation \ref{netCorDef}:
\begin{equation}
\rho_{XY}(k)={\frac{\mathbf{x}^{c}(k)^T\hat{\boldsymbol{\Sigma}}_{XY}^{(h)}\mathbf{y}^{c}(k)}{\sqrt{\mathbf{x}^{c}(k)^T\hat{\boldsymbol{\Sigma}}_{XX}^{(h)}\mathbf{x}^{c}(k)}\sqrt{\mathbf{y}^{c}(k)^T\hat{\boldsymbol{\Sigma}}_{YY}^{(h)}\mathbf{y}^{c}(k)}}}, \label{netCorDef}
\end{equation}
\noindent where $\hat{\boldsymbol{\Sigma}}_{XX}^{(h)}$, $\hat{\boldsymbol{\Sigma}}_{YY}^{(h)}$ and $\hat{\boldsymbol{\Sigma}}_{XY}^{(h)}$ are estimated from healthy rather than cancer samples in the methylation data set, according to equations \ref{covMats1} - \ref{covMats3},
\begin{equation}
\hat{\boldsymbol{\Sigma}}_{XX}^{(h)}=\frac{1}{n_h}{\sum_{k \in\text{healthy}}\left(\mathbf{x}(k)-\hat{\boldsymbol{\mu}}_X^{(h)}\right)\left(\mathbf{x}(k)-\hat{\boldsymbol{\mu}}_X^{(h)}\right)^T} \label{covMats1}
\end{equation}
\begin{equation}
\hat{\boldsymbol{\Sigma}}_{YY}^{(h)}=\frac{1}{n_h}{\sum_{k \in\text{healthy}}\left(\mathbf{y}(k)-\hat{\boldsymbol{\mu}}_Y^{(h)}\right)\left(\mathbf{y}(k)-\hat{\boldsymbol{\mu}}_Y^{(h)}\right)^T} \label{covMats2}
\end{equation}
\begin{equation}
\hat{\boldsymbol{\Sigma}}_{XY}^{(h)}=\frac{1}{n_h}{\sum_{k \in\text{healthy}}\left(\mathbf{x}(k)-\hat{\boldsymbol{\mu}}_X^{(h)}\right)\left(\mathbf{y}(k)-\hat{\boldsymbol{\mu}}_Y^{(h)}\right)^T}, \label{covMats3}
\end{equation}
where
\begin{equation*}
\hat{\boldsymbol{\mu}}_X^{(h)}=\frac{1}{n_h}\sum_{k \in\text{healthy}}\mathbf{x}(k),
\end{equation*}
\begin{equation*}
\hat{\boldsymbol{\mu}}_Y^{(h)}=\frac{1}{n_h}\sum_{k \in\text{healthy}}\mathbf{y}(k),
\end{equation*}
$n_h$ is the number of healthy samples in the data set, and $\mathbf{x}^{c}(k)$ and $\mathbf{y}^{c}(k)$ are the mean-centered methylation profiles $\mathbf{x}^{c}(k)=\mathbf{x}(k)-\hat{\boldsymbol{\mu}}_X^{(h)}$ and $\mathbf{y}^{c}(k)=\mathbf{y}(k)-\hat{\boldsymbol{\mu}}_Y^{(h)}$. The DNAm network interaction measure hence evaluates the extent to which, in an individual tumour sample, the combinations of the methylation-variables (i.e., loci) in genes $X$ and $Y$ explain each other, or covary, in the spaces determined by CCA on corresponding healthy samples; that is, the covariation in tumour sample $k$ between the methylation-variables in genes $X$ and $Y$ is assessed against typical healthy variability in these variables. When the DNA methylation network interaction measure $\rho_{XY}(k)$ is large (i.e., close to 1), the corresponding pair of genes explain each others' gene-regulatory behaviour (as reflected in their methylation profiles) well, or have otherwise well-correlated interactive or associative behaviour, for sample/patient $k$. Hence, $\rho_{XY}(k)$ measures (according to their DNA methylation profiles) the level of interaction or association between genes $X$ and $Y$ in tumour sample $k$, compared to typical interactions between these genes in healthy tissue.

\subsection{Prognostic Network Construction}\label{progNetConsSec}
Our proposed methodology for inference of network oncomarkers is based on a prognostic interaction network over $m$ genes. This network is represented by the $m\times m$ adjacency matrix $\mathbf{A}$, in which an edge is defined to to be present (i.e., $A_{ij}=1$) if and only if the corresponding pair of genes (nodes) are prognostic according to the DNA methylation network interaction measure of Section \ref{DNAmRhoSect}. Otherwise, we set $A_{ij}=0$. We note that $i$ and $j$ are now redefined compared to the last Section, so that they index genes rather than DNA methylation locations. This formulation will not be problematic, because all subsequent analysis is carried out at the level of genes rather than DNA methylation locations. To identify prognostic edges, we use the Cox proportional hazards model \citep{david1972regression} to calculate a Wald-statistic $z_{ij}$ for each of the $m\choose2$ pairs of genes in the network. The Wald statistic quantifies the strength of association of the DNA methylation network interaction measure $\rho_{ij}$ for the pair of genes $i$ and $j$ ($i=1,...,m$ and $j=1,...,m$) with patient survival outcome across patients $k$ ($k=1,...n$). We use a multivariate Cox model, adjusting these Wald statistics for clinical covariates, fitting this model separately to each pair of genes $(i,j)$. We adjust in this way in order to detect novel DNA methylation biomarkers which are independent of known prognostic clinical features.

The Wald statistic is asymptotically normally distributed with unit variance \citep{harrell2001regression}, and we can therefore model the distribution of our observed Wald statistics, $z_{ij}$, as a mixture of Gaussians. We have previously demonstrated the utility of mixture modelling to a related network inference problem \citep{bartlett2015network}, and a similar approach can be applied in this context. We model the $z_{ij}$ as a Gaussian mixture as follows:
\begin{align}
   z_{ij}\sim&\begin{cases}
    \mathcal{N}\left(\mu_{ij},\sigma^2\right), & \text{if}\quad A_{ij}=1,\\
    \mathcal{N}\left(0,\sigma^2\right), & \text{if}\quad A_{ij}=0,\label{nonZeroMixComp}
\end{cases}
\end{align}
where $\mathcal{N}\left(\mu_{ij},\sigma^2\right)$ is the normal distribution, and we enforce $\sigma^2=1$ in line with the asymptotic behaviour of the Wald statistic. We fit this mixture model to each observed statistic $z_{ij}$, and then infer whether, given $z_{ij}$, it is more likely that $\mu_{ij}=0$, or $\mu_{ij}\neq0$, leading to the estimates $\hat{A}_{ij}=0$ or $\hat{A}_{ij}=1$ respectively. We fit this model using the empirical Bayes procedure of \cite{johnstone2004needles}, defining a mixture prior distribution $f_\text{prior}\left(\mu_{ij}\right)$ over the $\mu_{ij}$ of equation $\ref{nonZeroMixComp}$:
\begin{equation}
f_\text{prior}\left(\mu_{ij}\right)=\left(1-w\right)\delta\left(\mu_{ij}\right)+w\gamma\left(\mu_{ij}\right),\label{mixPriorDefEq}
\end{equation}
where $w$ is the mixing parameter between the two components, which can also be interpreted as $w=\mathbb{E}\left[p\left(A_{ij}=1\right)\right]$, and $\gamma\left(\cdot\middle|a\right)$ is the Laplace probability density function,
\begin{equation*}
\gamma\left(\mu_{ij}\middle|a\right)=\frac{a}{2}\exp{\left(-a\left|\mu_{ij}\right|\right)},
\end{equation*}
where we use the standard value of $a=0.5$ \citep{johnstone2004needles}. Taking the mixture components to have Gaussian likelihoods, $f_\mathcal{N}\left(\cdot\middle|\mu_{ij},\sigma^2\right)$, as in equation \ref{nonZeroMixComp}, it follows from equation \ref{mixPriorDefEq} that the posterior density over the observed prognostic Wald statistic $z_{ij}$ is:
\begin{equation}
f_\text{posterior}\left(\mu_{ij}\middle|z_{ij}\right)=\frac{\left(1-w\right)\delta\left(\mu_{ij}\right)f_\mathcal{N}\left(z_{ij}\middle|0,\sigma^2\right)+w\gamma\left(\mu_{ij}\right)f_\mathcal{N}\left(z_{ij}\middle|\mu_{ij},\sigma^2\right)}{f_\text{marginal}\left(z_{ij}\right)},\label{mixPostDefEq}
\end{equation}
where the marginal density is:
\begin{equation}
f_\text{marginal}\left(z_{ij}\right)=(1-w)f_\mathcal{N}\left(z_{ij}\middle|0,\sigma^2\right)+wg\left(z_{ij}\right),\label{mixMargDefEQ}
\end{equation}
where $g\left(\mu_{ij}\right)$ is the convolution of the Laplace density with the standard normal density. If the Laplace distribution in the prior (equation \ref{mixPriorDefEq}) were replaced with a Gaussian, then the marginal distribution (equation \ref{mixMargDefEQ}) would be a mixture of Gaussians. However, as noted previously \citep{johnstone2004needles}, this empirical Bayes procedure requires a prior with tails that are exponential or heavier. Hence, we similarly use the Laplace rather than Gaussian prior which is a slight model misspecification. 

Although a separate model is fitted to each observed Wald statistic $z_{ij}$, a common weight $w_i$ is used for each gene/node $i$. We choose to do this, because estimating $w_i$ separately for each gene $i$ allows adaptation to a heterogenous degree distribution in $\mathbf{A}$, as follows. For a particular gene $i$, if the $z_{ij}$ are mostly close to zero, then $\hat{w}_i$ will be set low, which means that fewer edges ($A_{ij}=1$) will be detected; this hence corresponds to $i$ being a low-degree node. If for a different gene $i$ the $z_{ij}$ are generally further from zero, then $\hat{w}_i$ will be set high, which corresponds more edges being detected; this hence corresponds to $i$ being a high-degree node.

The estimate $\hat{w}_i$ is found as the value which maximises the marginal likelihood (equation \ref{margMLEdefEQ}) of the observed statistics $z_{ij}$ over all the pairwise comparisons of $i$ with $j$, $j\neq i$. This allows the model for each such pairwise comparison $(i,j)$ to `borrow strength' from all the other comparisons $(i,j')$, $j'\neq i$, $j'\neq j$: 
\begin{equation}
\hat{w}_i=\arg\max_w\sum_{j\neq i}\log\left\{(1-w)\phi\left(z_{ij}\right)+wg\left(z_{ij}\right)\right\}.\label{margMLEdefEQ}
\end{equation}
As in the original presentation of this methodology \citep{johnstone2004needles}, we use the posterior median to obtain the estimate $\hat{\mu}_{ij}$. Then we make a conservative estimate of $\mathbf{A}$ as follows:
\begin{align}
\hat{A}_{ij}=&1\quad\textrm{if}\quad\hat{\mu}_{ij}>0\enspace\textrm{and}\enspace\hat{\mu}_{ji}>0\quad\textrm{or}\quad\hat{\mu}_{ij}<0\enspace\textrm{and}\enspace\hat{\mu}_{ji}<0,\label{eBayesMixMLEdef}\\
\hat{A}_{ij}=&0\quad\textrm{otherwise}.\nonumber
\end{align}

\subsection{Community and Oncomarker Detection}\label{comDetSect}
Network nodes can be grouped together according to their propensity to interact with each other, for example groups of friends in a social network, or functional subnetwork modules in a biological network; this method is referred to as community detection \citep{girvan2002community,newman2004detecting}. We use community detection to naturally infer groups of genes in our constructed prognostic network. These groups of genes interact differently in cancer than in healthy tissue, in a way which is predictive of how advanced the disease is. We term these groups `network community oncomarkers'. Within a network community oncomarker the genes may interact with each other more (relative to healthy tissue) the more serious the disease is (as in Figure \ref{comGraphs}c), or they may interact with each other less the more serious the disease is, (as in Figure \ref{comGraphs}a). We carry out the task of community detection by fitting the degree-corrected stochastic blockmodel \citep{holland1983stochastic,bickel2009nonparametric}. We fit this model in an efficient way by regularised spectral clustering \citep{qin2013regularized}, calculating the optimum number of communities to divide the network into by the network histogram method \citep{olhede2014net}. Each community identified in this way represents a potential network community oncomarker.

For each network community oncomarker, we then calculate a prognostic score for each patient, by summarising the DNA methylation network interaction measure over this group of genes. This prognostic score can be used as a one-number summary of disease prognosis for that patient according to that network community oncomarker. The following points are important when calculating these summaries. Some gene-gene interactions will correspond to an increasingly negative DNA methylation network interaction measure $\rho_{ij}$ for worse patient prognosis. On the other hand, some gene-gene interactions will correspond to an increasingly positive $\rho_{ij}$ for worse prognosis. This means that care must be taken when summarising the network interaction measure across the network community oncomarker. Also, for the same amount of prognostic information conveyed, the magnitude of the changes in the network interaction measure may not be the same for each prognostic pairs of genes. To address these points, we combine the $\rho_{ij}$ across the prognostic pairs of genes of the network community after first multiplying them by the corresponding fitted Cox proportional hazards model coefficients $\hat{\theta}_{ij}$, obtained as described at the start of Section \ref{progNetConsSec}. Under the Cox proportional hazards model, the fitted model coefficient $\hat{\theta}_{ij}$ for a predictor $ij$ gives the log of the hazard ratio (HR) for that predictor in the model, that is, $\log\left(\text{HR}_{ij}\right)=\hat{\theta}_{ij}$. The hazard ratio is the scale-factor increase in probability of an event (e.g., death) occurring per unit time, relative to the baseline hazard (e.g., compared to a control group). Hence, these coefficients are interpretable in the same way, without scaling issues, across fitted models. This means that, for patient $k$, we can combine the DNA methylation network interaction measures over a network community oncomarker to generate a one-number prognostic score, as follows:
\begin{equation*}
\text{Score}_k=\sum_{i\in C,j\in C,i<j}\hat{A}_{ij}\hat{\theta}_{ij}\mathlarger{\rho}_{ij}(k),
\end{equation*}
where $C$ is the set of nodes in the network community oncomarker, $\hat{\mathbf{A}}$ is the inferred adjacency matrix, $\rho_{ij}(k)$ is the DNA methylation network interaction measure for genes/nodes $i$ and $j$ and patient $k$, and $\hat{\theta}_{ij}$ is the corresponding fitted Cox multivariate proportional-hazards model coefficient. Network edges/DNA methylation network interaction measures $\rho_{ij}$ which increase with poor prognosis (i.e., pairs of genes which interact more as the disease progresses, coloured green in Figure \ref{comGraphs}), will correspond to $\hat{\theta}_{ij} >0$. Hence, an increase in such a $\rho_{ij}$ will increase the prognostic score. Equivalently, network edges/DNA methylation network interaction measures $\rho_{ij}$ which decrease with poor prognosis (i.e., pairs of genes which interact less as the disease progresses, coloured red in Figure \ref{comGraphs}), will correspond to $\hat{\theta}_{ij} <0$. Hence, a decrease in such a $\rho_{ij}$ will also increase the prognostic score.

\subsection{An equivalent gene-expression interaction measure}
To examine further the hypothesis that the DNA methylation network interaction measure is a reflection of co-regulatory or co-regulated gene-expression patterns (among other genomic effects), we need an equivalent measure of gene-gene interaction or association in terms of gene expression. We can calculate such a measure, $\rho^\text{expr}_{XY}(k)$, for gene expression measurements $x^\text{expr}(k)$ and $y^\text{expr}(k)$ for the genes $X$ and $Y$ and patient $k$, as follows (equation \ref{geneExprIntMeasure}):
\begin{equation}
\rho^\text{expr}_{XY}(k)=\frac{\left(x^\text{expr}(k)-\hat{\mu}_{x^\text{expr}}^{(h)}\right)}{\hat{\sigma}_{x^\text{expr}}^{(h)}}\cdot\frac{\left(y^\text{expr}(k)-\hat{\mu}_{y^\text{expr}}^{(h)}\right)}{\hat{\sigma}_{y^\text{expr}}^{(h)}}\label{geneExprIntMeasure}
\end{equation}
where 
\begin{equation*}
\hat{\mu}_{x^\text{expr}}^{(h)}=\frac{1}{n_h}\sum_{k \in\text{healthy}}x^\text{expr}(k)\quad\text{and}\quad\hat{\mu}_{y^\text{expr}}^{(h)}=\frac{1}{n_h}\sum_{k \in\text{healthy}}y^\text{expr}(k),
\end{equation*}
\begin{equation*}
\left(\hat{\sigma}_{x^\text{expr}}^{(h)}\right)^2=\frac{1}{n_h}\sum_{k \in\text{healthy}}\left(x^\text{expr}(k)-\hat{\mu}_{x^\text{expr}}^{(h)}\right)^2
\end{equation*}
and
\begin{equation*}
\left(\hat{\sigma}_{y^\text{expr}}^{(h)}\right)^2=\frac{1}{n_h}\sum_{k \in\text{healthy}}\left(y^\text{expr}(k)-\hat{\mu}_{y^\text{expr}}^{(h)}\right)^2.
\end{equation*}
The intuition of equation \ref{geneExprIntMeasure} is that when the gene expression measurements $x^\text{expr}(k)$ and $y^\text{expr}(k)$ deviate \textit{in the same sample} from the corresponding healthy mean expression levels, this measure will be nonzero. When this occurs in the same samples as the DNA methylation network interaction measure $\rho_{XY}(k)$ is also nonzero, we will see a correlation between $\rho_{XY}(k)$ and $\rho^\text{expr}_{XY}$. These interaction measures for methylation and expression, $\rho_{XY}(k)$ and $\rho^\text{expr}_{XY}$, are equivalent because they both measure deviation from typical interactive behaviour in healthy/control samples.

\section{Examples}\label{resultsSect}
We present an example application of the methodology proposed in Section \ref{methodsSect} to a large publicly available breast cancer invasive carcinoma (BRCA) dataset downloaded from the Cancer Genome Atlas (TCGA). We downloaded an initial batch of DNA methylation data for tumour samples taken from 175 individuals (the training set), together with clinical data for these samples relating to patient survival outcome, and the covariates age, disease stage, and residual disease. These training data were used to detect potential network community oncomarkers. We then downloaded DNA methylation data for a further 528 tumour samples (the test set), together with data for the same clinical features: these independent samples were used to validate the potential network community oncomarkers. We also downloaded corresponding DNA methylation data for healthy breast tissue samples from 98 individuals to form a reference population of DNA methylation profiles for this analysis, and we downloaded gene expression data for 216 of the tumours for which DNA methylation data were also available. To proceed, we estimated from the training set the healthy population means, covariances and cross-covariances required to calculate the $\rho_{ij}$ ($i=1,...,m$ and $j=1,...,m$), as well as the corresponding log hazard ratios $\hat{\theta}_{ij}$ and adjacency matrix $\hat{\mathbf{A}}$. Additionally from the training data we estimated the communities in the adjacency matrix (including the number of communities) and prognostic score thresholds used to assign patients to better and worse prognostic groups. We then used these estimates to verify the prognostic ability of the methodology in the test set.

\begin{figure}
\centering\includegraphics[width=0.9\textwidth]{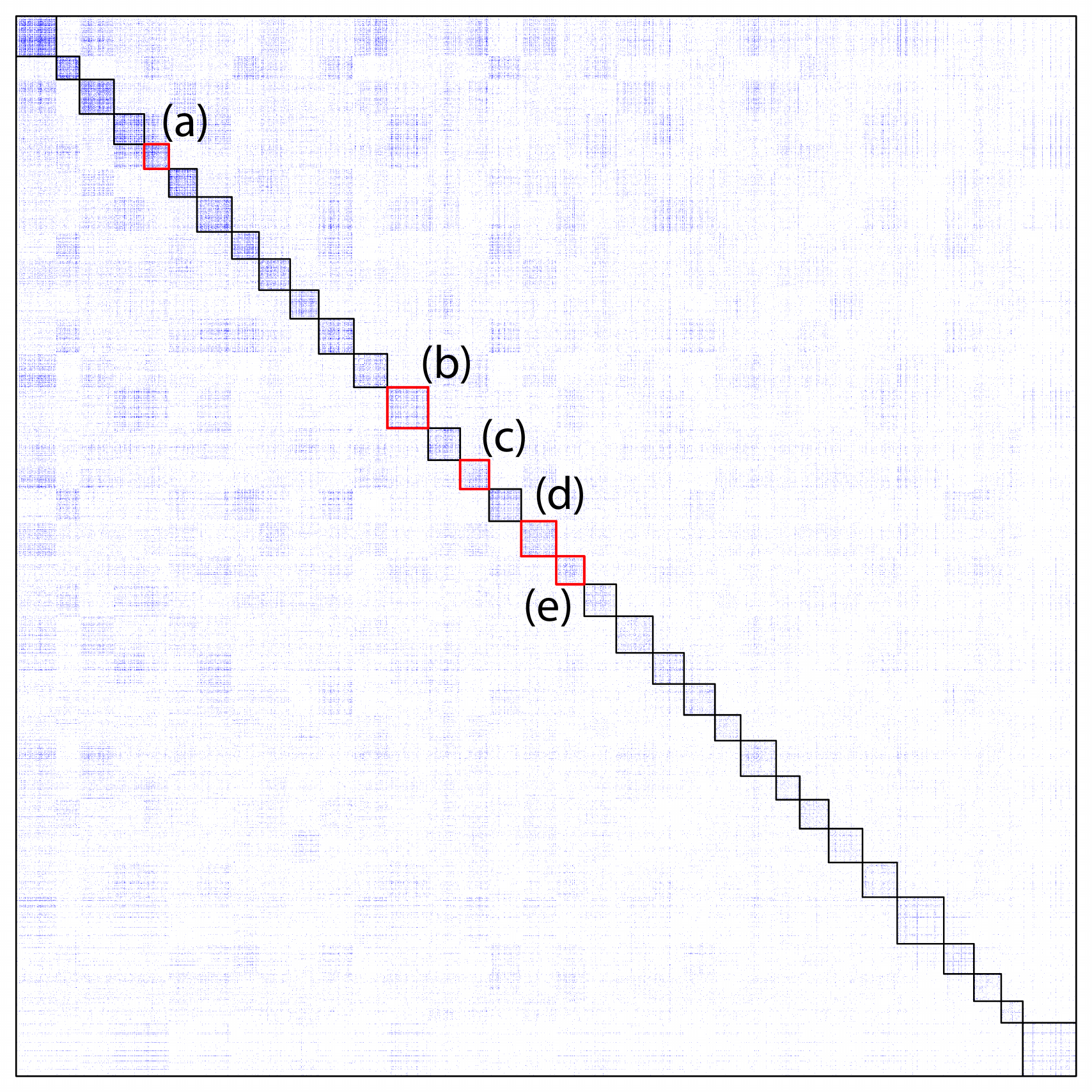}
\caption{The inferred prognostic adjacency matrix after community detection.} \label{adjMatHeat}
\vspace{-3ex}
\caption*{Entries in the adjacency matrix equal to 1 (representing a network edge) are coloured blue. Detected communities are outlined in black. The potential network community oncomarkers which are analysed further in Figures \ref{KMprognosis} - \ref{exprCorHists} and Tables \ref{progCoxRegTab} - \ref{validationCoxRegTab} and Tables S1 - S5 in the supplement are outlined in red, and labelled (a) - (e).}
\end{figure}

\begin{figure}
\vspace{-2ex}
\centering\includegraphics[width=0.9\textwidth]{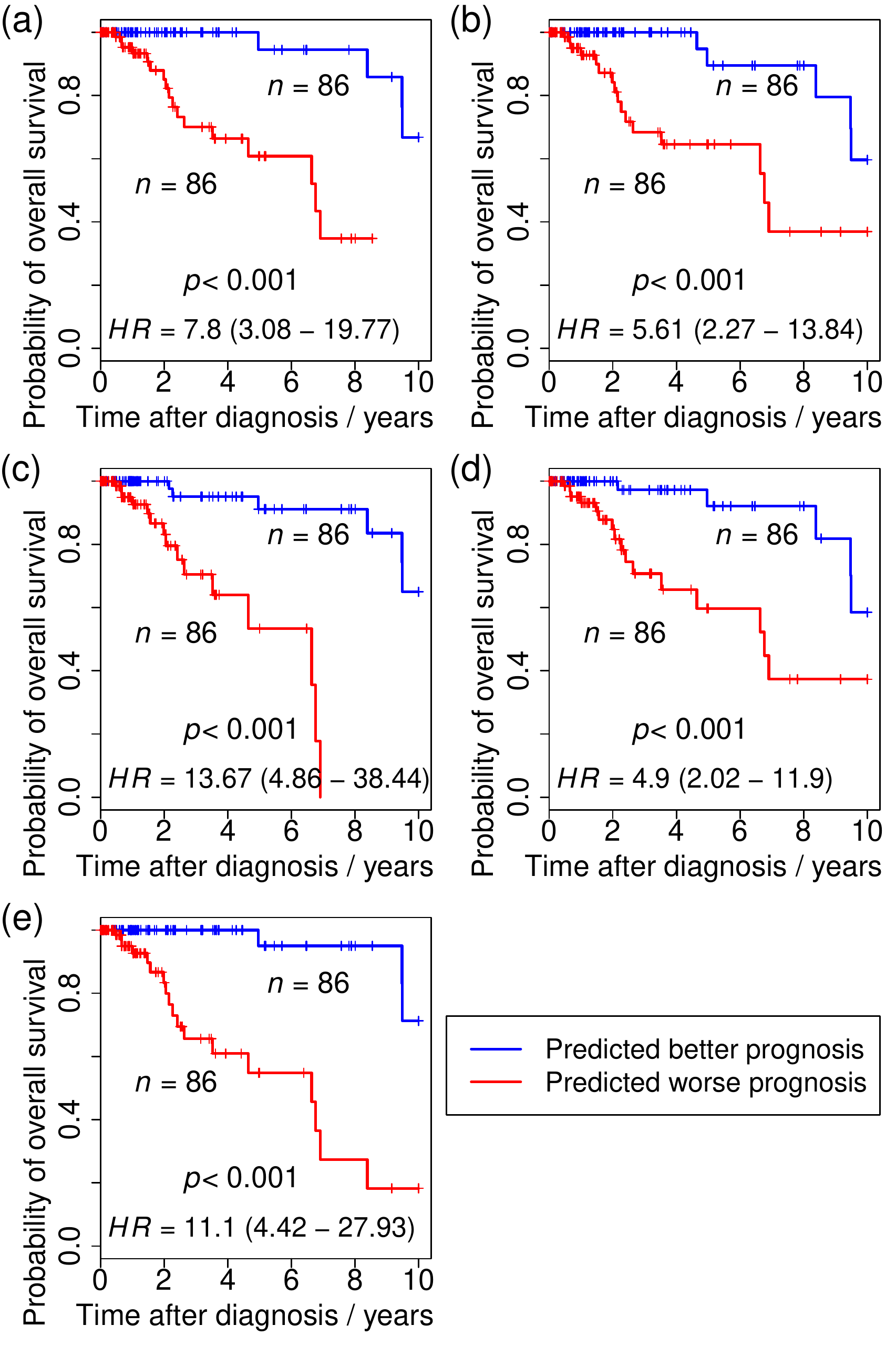}
\vspace{-2ex}
\caption{Network community oncomarkers: Kaplan-Meier plots for the training set.} \label{KMprognosis}
\vspace{-3ex}
\caption*{Comparison of survival curves for the patient groups defined by the prognostic score for each network community oncomarker. The groups are divided by the median prognostic score in the 175 samples of the training data set. The hazard ratio ($HR$) is displayed with 95\% C.I. in brackets, with the corresponding $p$-value calculated by univariate Cox regression. (a) - (e) indicate network community oncomarkers 1 - 5, as shown in Figure \ref{adjMatHeat}.}
\end{figure}

\begin{figure}
\vspace{-2ex}
\centering\includegraphics[width=0.9\textwidth]{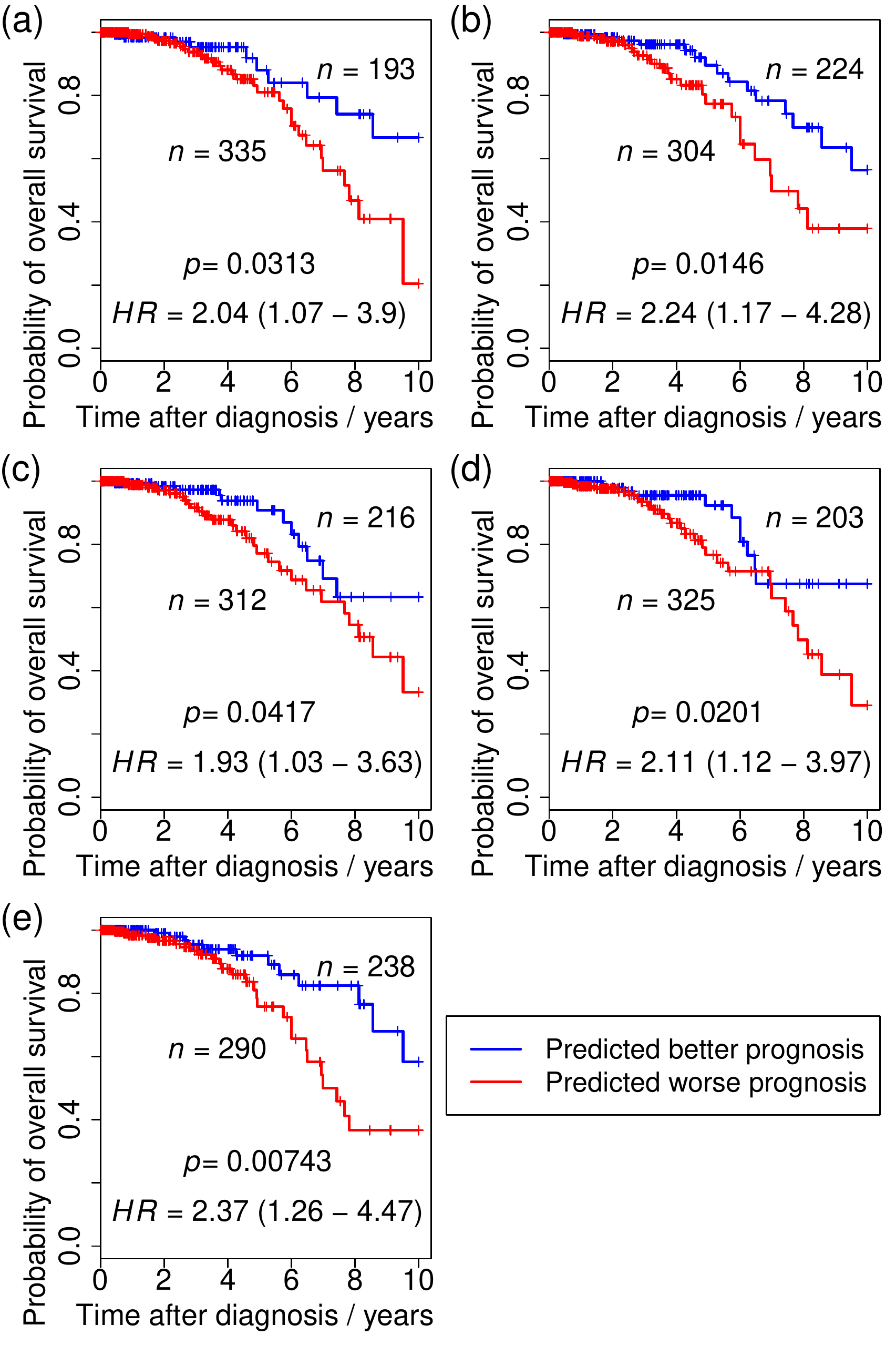}
\vspace{-2ex}
\caption{Network community oncomarkers: Kaplan-Meier plots for the test set.} \label{KMverify}
\vspace{-3ex}
\caption*{Comparison of survival curves for the patient groups defined by the prognostic score for each network community oncomarker. The groups are divided by the median prognostic score in the 175 samples of the training data set. The hazard ratio ($HR$) is displayed with 95\% C.I. in brackets, with the corresponding $p$-value calculated by univariate Cox regression. (a) - (e) indicate network community oncomarkers 1 - 4, as shown in Figure \ref{adjMatHeat}.}
\end{figure}

We first inferred the binary prognostic adjacency matrix $\hat{\mathbf{A}}$ for the 175 samples of the BRCA training data set according to the methods set out in Sections \ref{DNAmRhoSect} - \ref{progNetConsSec}. DNA methylation data were available for 14829 genes, and hence the number of nodes/genes $m$ in the inferred adjacency matrix $\hat{\mathbf{A}}$ is $m=14829$. The presence of an edge in $\hat{\mathbf{A}}$, that is, $\hat{A}_{ij}=1$, indicates that the interaction between genes $i$ and $j$ is associated with disease progression. The edge density of $\hat{\mathbf{A}}$ is $0.0035$, that is, $p(\hat{A}_{ij}=1)=0.0035$. We then extracted the connected component from this inferred network and carried out community detection on this connected component as described in Section \ref{comDetSect}. This resulted in 33 communities ranging from 116 to 285 nodes in size. The reduced adjacency matrix relating to these communities [with $m=5668$ and $p(\hat{A}_{ij}=1)=0.023$] is shown in Figure $\ref{adjMatHeat}$. We note that the stochastic blockmodel, fitted in this way via spectral clustering, does not provide any uncertainty as to the inferred community assignments: if this is desired, then mixed-membership stochastic blockmodels are available as an alternative \citep{airoldi2008mixed}. In the analysis we present here, uncertainties arising from these inferred community assignments are considered in the subsequent analyses (Figures \ref{KMprognosis} and \ref{KMverify} and Tables \ref{progCoxRegTab} and \ref{validationCoxRegTab}).

We validated each of the 33 potential network community oncomarkers in the 528 independent tumour samples of the test/validation set. We note that these 528 samples were not used in any way to identify the 33 potential network community oncomarkers shown in Figure \ref{adjMatHeat}. Hence in this validation each of these 528 patients were classified individually according to prognosis without reference to the other validation samples. This means that comparing these prognostic classifications assigned to the validation samples is a true test of prognostic ability of the network community oncomarkers. To carry out the validation, we calculated the prognostic score for the 528 independent/unseen samples of the test set, based on the inferred prognostic adjacency matrix $\hat{\mathbf{A}}$ and the fitted Cox multivariate proportional hazards model coefficients $\hat{\boldsymbol{\theta}}$ obtained from the initial 175 samples of the training set. Using this trained model, we calculated one prognostic score for each potential network community oncomarker for each of the 528 unseen test-set samples. We then tested the prognostic score, for each potential network community oncomarker, for significant prediction of patient survival outcome in these 528 unseen test-set samples. The five potential network community oncomarkers which validated in this way with the highest level of significance are outlined in red in Figure $\ref{adjMatHeat}$. The results of univariate and multivariate Cox regression for these five best network community oncomarkers are shown in Figures \ref{KMprognosis} and \ref{KMverify}, and in Tables \ref{progCoxRegTab} and \ref{validationCoxRegTab}, for the training and test sets respectively. Plots equivalent to Figures \ref{KMprognosis} and \ref{KMverify} for all 33 detected network communities appear in Supplementary Figures S1-S2. For the multivariate analysis, samples with missing data for any of the clinical covariates were removed, leaving 172 and 396 samples for the training and test sets respectively. We note that, as would be expected, the level of significance in the training set (to which the model was fitted, Figure \ref{KMprognosis} and Table\ref{progCoxRegTab}), is much higher than in the test set (Figure \ref{KMverify} and Table \ref{validationCoxRegTab}). 

\begin{table}
\centering
\begin{tabular}{|r|l|l|l|}
          \hline
         & HR (95\%CI) & \textit{p} & \textit{n} \\ 
          \hline
Prognostic Score & 77.1 (10.5-567) & $<$0.001 & 172 \\ 
  Age & 1.79 (0.66-4.84) & 0.249 & 172 \\ 
  Residual Disease & 15.4 (4.68-50.9) & $<$0.001 & 172 \\ 
  Stage & 2.85 (0.96-8.46) & 0.060 & 172 \\ 
           \hline
\multicolumn{4}{l}{(a) Network community oncomarker 1.}\\
\multicolumn{4}{l}{ }\\
      \hline
     & HR (95\%CI) & \textit{p} & \textit{n} \\ 
      \hline
  Prognostic Score & 51.3 (8.35-315) & $<$0.001 & 172 \\ 
  Age & 1.42 (0.48-4.23) & 0.53 & 172 \\ 
  Residual Disease & 30.4 (5.82-158) & $<$0.001 & 172 \\ 
  Stage & 1.95 (0.68-5.54) & 0.212 & 172 \\ 
           \hline
\multicolumn{4}{l}{(b) Network community oncomarker 2.}\\
\multicolumn{4}{l}{ }\\
      \hline
               & HR (95\%CI) & \textit{p} & \textit{n} \\ 
          \hline
  Prognostic Score & 50.1 (9.77-256) & $<$0.001 & 172 \\ 
  Age & 2.16 (0.81-5.8) & 0.125 & 172 \\ 
  Residual Disease & 13.3 (4.54-39.1) & $<$0.001 & 172 \\ 
  Stage & 2.41 (0.81-7.18) & 0.114 & 172 \\ 
           \hline
\multicolumn{4}{l}{(c) Network community oncomarker 3.}\\
\multicolumn{4}{l}{ }\\
      \hline
     & HR (95\%CI) & \textit{p} & \textit{n} \\ 
      \hline
    Prognostic Score & 22.7 (5.52-93.1) & $<$0.001 & 172 \\ 
  Age & 3.49 (1.3-9.42) & 0.0135 & 172 \\ 
  Residual Disease & 16.3 (5.24-50.7) & $<$0.001 & 172 \\ 
  Stage & 1.05 (0.38-2.91) & 0.928 & 172 \\ 
           \hline
\multicolumn{4}{l}{(d) Network community oncomarker 4.}\\
\multicolumn{4}{l}{ }\\
      \hline
         & HR (95\%CI) & \textit{p} & \textit{n} \\ 
          \hline
  Prognostic Score & 46.0 (8.17-259) & $<$0.001 & 172 \\ 
  Age & 2.91 (1-8.44) & 0.0493 & 172 \\ 
  Residual Disease & 7.04 (2.68-18.5) & $<$0.001 & 172 \\ 
  Stage & 3.74 (1.23-11.4) & 0.02 & 172 \\ 
           \hline
\multicolumn{4}{l}{(e) Network community oncomarker 5.}\\
\multicolumn{4}{l}{ }\\
               \end{tabular}
\caption{Network community oncomarkers - training set prognosis.\\ Multivariate Cox regression was used to test significance of the prognostic scores obtained from the network community oncomarkers. (a) - (e) indicate network community oncomarkers 1 - 5, as shown in Figure \ref{adjMatHeat}.}\label{progCoxRegTab}
\end{table}

\begin{table}
\centering
\begin{tabular}{|r|l|l|l|}
          \hline
         & HR (95\%CI) & \textit{p} & \textit{n} \\ 
          \hline
Prognostic Score & 4.89 (1.65-14.5) & 0.00429 & 396 \\ 
  Age & 3.52 (1.46-8.49) & 0.00513 & 396 \\ 
  Residual Disease & 12.5 (5.32-29.3) & $<$0.001 & 396 \\ 
  Stage & 1.62 (0.66-4) & 0.294 & 396 \\ 
           \hline
\multicolumn{4}{l}{(a) Network community oncomarker 1.}\\
\multicolumn{4}{l}{ }\\
      \hline
     & HR (95\%CI) & \textit{p} & \textit{n} \\ 
      \hline
  Prognostic Score & 5.07 (1.81-14.1) & 0.00195 & 396 \\ 
  Age & 3.67 (1.49-9.03) & 0.00458 & 396 \\ 
  Residual Disease & 8.72 (3.78-20.1) & $<$0.001 & 396 \\ 
  Stage & 1.47 (0.6-3.61) & 0.406 & 396 \\ 
           \hline
\multicolumn{4}{l}{(b) Network community oncomarker 2.}\\
\multicolumn{4}{l}{ }\\
      \hline
         & HR (95\%CI) & \textit{p} & \textit{n} \\ 
          \hline
  Prognostic Score & 2.63 (1.01-6.89) & 0.0484 & 396 \\ 
  Age & 2.07 (0.86-5) & 0.106 & 396 \\ 
  Residual Disease & 11.3 (4.97-25.5) & $<$0.001 & 396 \\ 
  Stage & 2.04 (0.76-5.45) & 0.157 & 396 \\ 
           \hline
\multicolumn{4}{l}{(c) Network community oncomarker 3.}\\
\multicolumn{4}{l}{ }\\
      \hline
     & HR (95\%CI) & \textit{p} & \textit{n} \\ 
      \hline
    Prognostic Score & 4.92 (1.8-13.5) & 0.00189 & 396 \\ 
  Age & 1.91 (0.78-4.69) & 0.159 & 396 \\ 
  Residual Disease & 17.2 (6.76-43.9) & $<$0.001 & 396 \\ 
  Stage & 0.92 (0.34-2.48) & 0.871 & 396 \\ 
           \hline
\multicolumn{4}{l}{(d) Network community oncomarker 4.}\\
\multicolumn{4}{l}{ }\\
      \hline
         & HR (95\%CI) & \textit{p} & \textit{n} \\ 
          \hline
  Prognostic Score & 2.5 (0.94-6.65) & 0.0668 & 396 \\ 
  Age & 2.23 (0.94-5.27) & 0.0677 & 396 \\ 
  Residual Disease & 8.17 (3.47-19.3) & $<$0.001 & 396 \\ 
  Stage & 1.59 (0.64-3.95) & 0.321 & 396 \\ 
           \hline
\multicolumn{4}{l}{(e) Network community oncomarker 5.}\\
\multicolumn{4}{l}{ }\\
               \end{tabular}
\caption{Network community oncomarkers - test/validation set prognosis.\\ Multivariate Cox regression was used to test significance of the prognostic scores obtained from the network community oncomarkers. (a) - (e) indicate network community oncomarkers 1 - 5, as shown in Figure \ref{adjMatHeat}.}\label{validationCoxRegTab}
\end{table}

\begin{figure}[ht!]
\centering\includegraphics[width=0.9\textwidth]{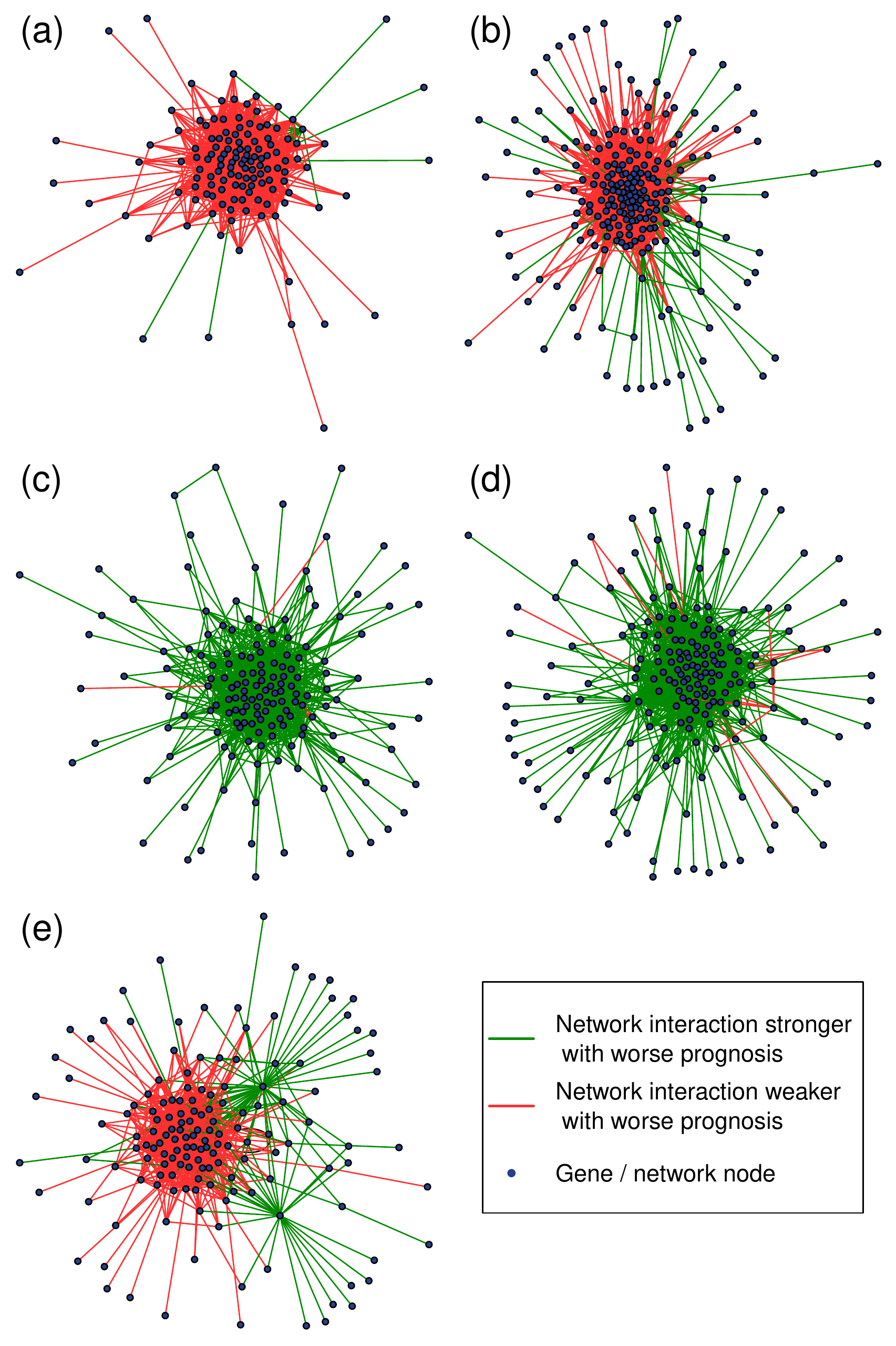}
\caption{Detected network community oncomarkers.} \label{comGraphs}
\vspace{-3ex}
\caption*{(a) - (e) indicate network community oncomarkers 1 - 5, as shown in Figure \ref{adjMatHeat}.}
\end{figure}

Figure \ref{comGraphs} shows the five network community oncomarkers which validated most significantly. Green edges indicate gene-gene interactions which become stronger with disease progression. Red edges indicate interactions which become weaker with disease progression. Hence, the network community oncomarkers of Figures \ref{comGraphs}a and \ref{comGraphs}b can be considered to be functional subnetwork modules which become less active as the cancer progresses (comprised of 99\% and 96\% red edges, respectively). On the other hand, Figures \ref{comGraphs}c and \ref{comGraphs}d can be considered to be functional subnetwork modules which become more active as the cancer progresses (both comprised of 99\% green edges). Then the network community oncomarker of Figure \ref{comGraphs}e contains a mixture of these effects (comprised of 87\% red and 13\% green edges). However, each of these network community oncomarkers represents a functional subnetwork module which is rewired in a way which is advantageous for the cancer, in favour of proliferation, and against cell death and immune function. The genes/nodes of these network community oncomarkers are shown in Tables S1 - S5 in the supplement; they list many genes related to cell proliferation (e.g., \textit{CDKL1}, \textit{NKAPL}, \textit{MAPK6}), developmental processes (e.g., \textit{HOXD10}, \textit{HOXB9}, \textit{HOXC10}, \textit{HOXA13}, \textit{HOXC12}, \textit{HOXD13}), and immune function (e.g., \textit{VSIG2}, \textit{IL36B}, \textit{RBPJ}).

\begin{figure}[ht!]
\centering\includegraphics[width=0.9\textwidth]{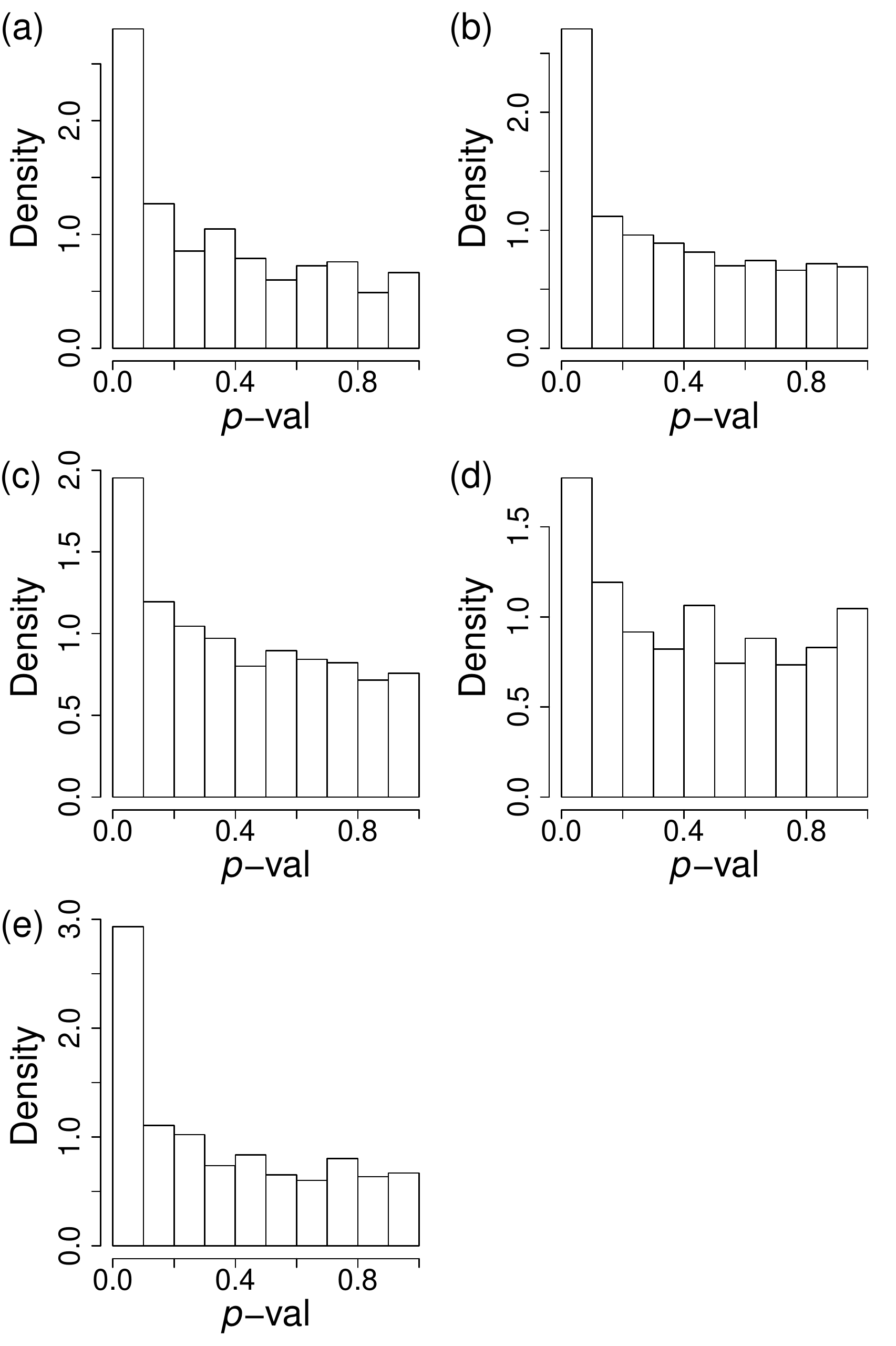}
\caption{Correlation of DNA methylation with gene expression for the network community oncomarkers.} \label{exprCorHists}
\vspace{-3ex}
\caption*{(a) - (e) indicate network community oncomarkers 1 - 5, as shown in Figure \ref{adjMatHeat}.}
\vspace{-3ex}
\end{figure}

We hypothesise that the DNA methylation network interaction measure is a reflection of co-regulatory or co-regulated gene-expression patterns, among other genomic effects. We tested this hypothesis by comparing the DNA methylation network interaction measure $\rho_{XY}$ for a pair of genes $XY$ (equation \ref{netCorDef}) with an equivalent measure of interactive behaviour of these genes in terms of their expression levels, $\rho^\text{expr}_{XY}$ (equation \ref{geneExprIntMeasure}). Correlation test $p$-values for the comparison between $\rho_{XY}$ and $\rho^\text{expr}_{XY}$ appear in Figure \ref{exprCorHists}. It is clear that in these histograms, there is a concentration of significant $p$-values close to zero, indicating a departure from the null hypothesis uniform distribution, and demonstrating an association between $\rho_{XY}$ and $\rho^\text{expr}_{XY}$ for many of the edges/interactions of each network community oncomarker. However, there are also many nonsignificant $p$-values visible in these histograms, indicating that there are other genomic interactive effects present which cannot be explained in terms of gene expression (as assessed by mRNA levels) alone. Such effects are expected to include the influence of alternatively spliced products or isoforms \citep{jones2012functions} and the interaction between noncoding transcripts and the epigenome \citep{lai2014long}.

\section{Discussion}
In this paper, we have proposed methodology to detect cancer biomarkers based on the epigenomic pattern DNA methylation. This methodology builds on a previously proposed measure of pairwise interaction between genes, based on the epigenomic gene-regulatory pattern DNA methylation \citep{bartlett2014dna}. Based on this DNA methylation network interaction measure, the methodology we describe in this paper allows inference of prognostic genomic networks, and identification of prognostic biomarkers from such networks using community detection methodology. Community detection has previously proved powerful as well realistic in a range of fields, including social as well as biological networks \citep{girvan2002community}. In the context of genomic networks, such modular groups of genes are known to correspond to specific physiological functions \citep{shen2002network}. The modular prognostic biomarkers which we detect are termed `network community oncomarkers'; they are groups of nodes/genes among which there is a high density of prognostic genomic interactive or associative behaviour. We have demonstrated that within these communities, the DNA methylation network interaction measure is highly associated with co-regulatory behaviour linked to gene expression (at the mRNA level), giving functional relevance to the findings. However, there are also likely to be a range of genomic interactive effects present which are measured by the DNA methylation network interaction measure but which are not reflected in mRNA levels. Our proposed methodology also allows a one-number prognostic score for a network community oncomarker to be calculated for each patient/sample: this prognostic score is a measure of disease progression in that patient.

Our proposed methodology uses mixture modelling to infer network structure from prognostic association between genes, and draws on practical approaches to community detection to obtain oncomarkers from this prognostic network. Mixture modelling has previously been shown to be an effective approach to the related problem of clustering in networks \citep{vu2013model}. This suggests that more general methodology could be developed here, in which network and community inference are both carried out simultaneously by model fitting. Network inference has also been carried out previously using multiple node attributes in cell biological data \citep{katenka2012inference}, and those findings could be used as a basis upon which data from other genomic sources could be integrated into the methodology proposed here. Genes also frequently carry out multiple roles in different biological contexts and hence may be involved in more than one functional subnetwork module within a genomic network. Work has been carried out on overlapping stochastic blockmodels \citep{latouche2011overlapping}, and hence this would be a natural context in which to develop an application for such methodology.

The field of epigenomics is progressing fast and promises many new insights in the near future into unexplained or undiscovered genomic phenomena, for example relating to the so-called `dark matter' of the genome \citep{venters2013genomic}. Epigenomics is also expected to provide new understanding of the mechanisms of disease progression. The discovery that some genomic loci gain or lose methylation in ways which may be unique to cancer suggests that understanding changes in DNA methylation machinery may be essential to understanding oncogenesis \citep{xie2013epigenomic}. The field of network science is also advancing rapidly. Networks are an efficient way to represent and analyse large numbers of variables, which is particularly relevant in modern, large-scale genomic studies. Networks of interactions are a natural way to represent and analyse genomic interactions, associations and processes. Therefore, the study of genomic and epigenomic networks promises to be productive over the coming years for the fields of biology, medicine, and statistics.

\section{Datasets}
DNA methylation (DNAm) data from breast cancer invasive carcinoma (BRCA) tumour samples, collected via the Illumina Infinium HumanMethylation450 platform, were downloaded from The Cancer Genome Atlas (TCGA) project \citep{hampton2006cancer,bonetta2006genome,collins2007mapping} at level 3. These data were preprocessed by first removing probes with nonunique mappings and which map to SNPs (as identified in the TCGA level 3 data); probes mapping to sex chromosomes were also removed; in total 98384 probes were removed in this way from all data sets. After removal of these probes, 270985 probes with known gene annotations remained. Probes were then removed if they had less than 95\% coverage across samples; probe values were also replaced if they had corresponding detection $p$-value greater than 5\%, by KNN ($k$ nearest neighbour) imputation ($k=5$). The loci of analysed CpGs were mapped to genes based on annotation information for the Illumina Infinium platform obtained from the \textit{R} / \textit{Bioconductor} package \textquoteleft IlluminaHumanMethylation450k\textquoteright. The data were also checked for batch effects by hierarchical clustering and correlation of the significant principle components with phenotype and batch: no significant batch effects (which would warrant further correction) were found. We downloaded DNA methylation data for tumour samples from 175 samples/individuals, from TCGA in July 2013, with clinical data available for patient survival outcome, and the clinical covariates age, disease stage, and residual disease. At the same time, we also downloaded corresponding DNA methylation data for healthy tissue for 98 individuals. These data were used to detect potential network community oncomarkers. We then downloaded DNA methylation data for a further 528 tumour samples from TCGA in September 2014, with data for the same clinical features available. These independent samples were used to validate the potential network community oncomarkers. At this time we also downloaded gene expression data from TCGA at level 3, for 216 of the tumours for which we also obtained DNA methylation data.

\bibliography{../../references}

\begin{thebibliography}{59}
% BibTex style file: imsart-nameyear.bst, 2013-01-28
% Default style options (sort=1,type=nameyear).
% Used options (sort=1,type=nameyear).

\bibitem[\protect\citeauthoryear{Airoldi et~al.}{2008}]{airoldi2008mixed}
\begin{barticle}[author]
\bauthor{\bsnm{Airoldi},~\bfnm{Edoardo~M}\binits{E.~M.}},
  \bauthor{\bsnm{Blei},~\bfnm{David~M}\binits{D.~M.}},
  \bauthor{\bsnm{Fienberg},~\bfnm{Stephen~E}\binits{S.~E.}} \AND
  \bauthor{\bsnm{Xing},~\bfnm{Eric~P}\binits{E.~P.}}
(\byear{2008}).
\btitle{Mixed membership stochastic blockmodels}.
\bjournal{Journal of Machine Learning Research}
\bvolume{9}
\bpages{1981--2014}.
\end{barticle}
\endbibitem

\bibitem[\protect\citeauthoryear{Barab{\'a}si and
  Oltvai}{2004}]{barabasi2004network}
\begin{barticle}[author]
\bauthor{\bsnm{Barab{\'a}si},~\bfnm{Albert-L{\'a}szl{\'o}}\binits{A.-L.}} \AND
  \bauthor{\bsnm{Oltvai},~\bfnm{Zoltan~N}\binits{Z.~N.}}
(\byear{2004}).
\btitle{Network biology: understanding the cell's functional organization}.
\bjournal{Nature Reviews Genetics}
\bvolume{5}
\bpages{101--113}.
\end{barticle}
\endbibitem

\bibitem[\protect\citeauthoryear{Bartlett}{2015}]{bartlett2015network}
\begin{barticle}[author]
\bauthor{\bsnm{Bartlett},~\bfnm{Thomas~E}\binits{T.~E.}}
(\byear{2015}).
\btitle{Network inference and community detection, based on covariance
  matrices, correlations and test statistics from arbitrary distributions}.
\bjournal{arXiv preprint arXiv:1506.04928}.
\end{barticle}
\endbibitem

\bibitem[\protect\citeauthoryear{Bartlett, Olhede and
  Zaikin}{2014}]{bartlett2014dna}
\begin{barticle}[author]
\bauthor{\bsnm{Bartlett},~\bfnm{Thomas~E}\binits{T.~E.}},
  \bauthor{\bsnm{Olhede},~\bfnm{Sofia~C}\binits{S.~C.}} \AND
  \bauthor{\bsnm{Zaikin},~\bfnm{Alexey}\binits{A.}}
(\byear{2014}).
\btitle{A {DNA} Methylation Network Interaction Measure, and Detection of
  Network Oncomarkers}.
\bjournal{PloS One}
\bvolume{9}
\bpages{e84573}.
\end{barticle}
\endbibitem

\bibitem[\protect\citeauthoryear{Bartlett
  et~al.}{2013}]{bartlett2013corruption}
\begin{barticle}[author]
\bauthor{\bsnm{Bartlett},~\bfnm{Thomas~E}\binits{T.~E.}},
  \bauthor{\bsnm{Zaikin},~\bfnm{Alexey}\binits{A.}},
  \bauthor{\bsnm{Olhede},~\bfnm{Sofia~C}\binits{S.~C.}},
  \bauthor{\bsnm{West},~\bfnm{James}\binits{J.}},
  \bauthor{\bsnm{Teschendorff},~\bfnm{Andrew~E}\binits{A.~E.}} \AND
  \bauthor{\bsnm{Widschwendter},~\bfnm{Martin}\binits{M.}}
(\byear{2013}).
\btitle{Corruption of the Intra-Gene {DNA} Methylation Architecture Is a
  Hallmark of Cancer}.
\bjournal{PloS One}
\bvolume{8}
\bpages{e68285}.
\end{barticle}
\endbibitem

\bibitem[\protect\citeauthoryear{Beguerisse-D{\'\i}az
  et~al.}{2014}]{beguerisse2014interest}
\begin{barticle}[author]
\bauthor{\bsnm{Beguerisse-D{\'\i}az},~\bfnm{Mariano}\binits{M.}},
  \bauthor{\bsnm{Gardu{\~n}o-Hern{\'a}ndez},~\bfnm{Guillermo}\binits{G.}},
  \bauthor{\bsnm{Vangelov},~\bfnm{Borislav}\binits{B.}},
  \bauthor{\bsnm{Yaliraki},~\bfnm{Sophia~N}\binits{S.~N.}} \AND
  \bauthor{\bsnm{Barahona},~\bfnm{Mauricio}\binits{M.}}
(\byear{2014}).
\btitle{Interest communities and flow roles in directed networks: the Twitter
  network of the UK riots}.
\bjournal{Journal of The Royal Society Interface}
\bvolume{11}
\bpages{20140940}.
\end{barticle}
\endbibitem

\bibitem[\protect\citeauthoryear{Bhagat et~al.}{2012}]{bhagat2012aberrant}
\begin{barticle}[author]
\bauthor{\bsnm{Bhagat},~\bfnm{Rahul}\binits{R.}},
  \bauthor{\bsnm{Chadaga},~\bfnm{Shilpa}\binits{S.}},
  \bauthor{\bsnm{Premalata},~\bfnm{CS}\binits{C.}},
  \bauthor{\bsnm{Ramesh},~\bfnm{G}\binits{G.}},
  \bauthor{\bsnm{Ramesh},~\bfnm{C}\binits{C.}},
  \bauthor{\bsnm{Pallavi},~\bfnm{VR}\binits{V.}} \AND
  \bauthor{\bsnm{Krishnamoorthy},~\bfnm{Lakshmi}\binits{L.}}
(\byear{2012}).
\btitle{Aberrant promoter methylation of the RASSF1A and APC genes in
  epithelial ovarian carcinoma development}.
\bjournal{Cellular Oncology}
\bvolume{35}
\bpages{473--479}.
\end{barticle}
\endbibitem

\bibitem[\protect\citeauthoryear{Bickel and
  Chen}{2009}]{bickel2009nonparametric}
\begin{barticle}[author]
\bauthor{\bsnm{Bickel},~\bfnm{Peter~J}\binits{P.~J.}} \AND
  \bauthor{\bsnm{Chen},~\bfnm{Aiyou}\binits{A.}}
(\byear{2009}).
\btitle{A nonparametric view of network models and Newman--Girvan and other
  modularities}.
\bjournal{Proceedings of the National Academy of Sciences}
\bvolume{106}
\bpages{21068--21073}.
\end{barticle}
\endbibitem

\bibitem[\protect\citeauthoryear{Bonetta}{2006}]{bonetta2006genome}
\begin{barticle}[author]
\bauthor{\bsnm{Bonetta},~\bfnm{Laura}\binits{L.}}
(\byear{2006}).
\btitle{Genome sequencing in the fast lane}.
\bjournal{Nature Methods}
\bvolume{3}
\bpages{141}.
\end{barticle}
\endbibitem

\bibitem[\protect\citeauthoryear{Brocks et~al.}{2014}]{brocks2014intratumor}
\begin{barticle}[author]
\bauthor{\bsnm{Brocks},~\bfnm{David}\binits{D.}},
  \bauthor{\bsnm{Assenov},~\bfnm{Yassen}\binits{Y.}},
  \bauthor{\bsnm{Minner},~\bfnm{Sarah}\binits{S.}},
  \bauthor{\bsnm{Bogatyrova},~\bfnm{Olga}\binits{O.}},
  \bauthor{\bsnm{Simon},~\bfnm{Ronald}\binits{R.}},
  \bauthor{\bsnm{Koop},~\bfnm{Christina}\binits{C.}},
  \bauthor{\bsnm{Oakes},~\bfnm{Christopher}\binits{C.}},
  \bauthor{\bsnm{Zucknick},~\bfnm{Manuela}\binits{M.}},
  \bauthor{\bsnm{Lipka},~\bfnm{Daniel~Bernhard}\binits{D.~B.}},
  \bauthor{\bsnm{Weischenfeldt},~\bfnm{Joachim}\binits{J.}} \betal{et~al.}
(\byear{2014}).
\btitle{Intratumor {DNA} methylation heterogeneity reflects clonal evolution in
  aggressive prostate cancer}.
\bjournal{Cell Reports}
\bvolume{8}
\bpages{798--806}.
\end{barticle}
\endbibitem

\bibitem[\protect\citeauthoryear{Christensen
  et~al.}{2009}]{christensen2009epigenetic}
\begin{barticle}[author]
\bauthor{\bsnm{Christensen},~\bfnm{Brock~C}\binits{B.~C.}},
  \bauthor{\bsnm{Houseman},~\bfnm{E~Andres}\binits{E.~A.}},
  \bauthor{\bsnm{Marsit},~\bfnm{Carmen~J}\binits{C.~J.}},
  \bauthor{\bsnm{Zheng},~\bfnm{Shichun}\binits{S.}},
  \bauthor{\bsnm{Wrensch},~\bfnm{Margaret~R}\binits{M.~R.}},
  \bauthor{\bsnm{Wiemels},~\bfnm{Joseph~L}\binits{J.~L.}},
  \bauthor{\bsnm{Nelson},~\bfnm{Heather~H}\binits{H.~H.}},
  \bauthor{\bsnm{Karagas},~\bfnm{Margaret~R}\binits{M.~R.}},
  \bauthor{\bsnm{Padbury},~\bfnm{James~F}\binits{J.~F.}},
  \bauthor{\bsnm{Bueno},~\bfnm{Raphael}\binits{R.}} \betal{et~al.}
(\byear{2009}).
\btitle{Aging and environmental exposures alter tissue-specific {DNA}
  methylation dependent upon {CpG} island context}.
\bjournal{PLoS Genetics}
\bvolume{5}
\bpages{e1000602}.
\end{barticle}
\endbibitem

\bibitem[\protect\citeauthoryear{Clune, Mouret and
  Lipson}{2013}]{clune2013evolutionary}
\begin{barticle}[author]
\bauthor{\bsnm{Clune},~\bfnm{Jeff}\binits{J.}},
  \bauthor{\bsnm{Mouret},~\bfnm{Jean-Baptiste}\binits{J.-B.}} \AND
  \bauthor{\bsnm{Lipson},~\bfnm{Hod}\binits{H.}}
(\byear{2013}).
\btitle{The evolutionary origins of modularity}.
\bjournal{Proceedings of the Royal Society of London B: Biological Sciences}
\bvolume{280}
\bpages{20122863}.
\end{barticle}
\endbibitem

\bibitem[\protect\citeauthoryear{Collins and Barker}{2007}]{collins2007mapping}
\begin{barticle}[author]
\bauthor{\bsnm{Collins},~\bfnm{F.}\binits{F.}} \AND
  \bauthor{\bsnm{Barker},~\bfnm{A.}\binits{A.}}
(\byear{2007}).
\btitle{Mapping the cancer genome}.
\bjournal{Scientific American Magazine}
\bvolume{296}
\bpages{50--57}.
\end{barticle}
\endbibitem

\bibitem[\protect\citeauthoryear{Cooney}{2007}]{cooney2007epigenetics}
\begin{barticle}[author]
\bauthor{\bsnm{Cooney},~\bfnm{Craig~A}\binits{C.~A.}}
(\byear{2007}).
\btitle{Epigenetics-{DNA}-based mirror of our environment?}
\bjournal{Disease Markers}
\bvolume{23}
\bpages{121--137}.
\end{barticle}
\endbibitem

\bibitem[\protect\citeauthoryear{Cox}{1972}]{david1972regression}
\begin{barticle}[author]
\bauthor{\bsnm{Cox},~\bfnm{David~R}\binits{D.~R.}}
(\byear{1972}).
\btitle{Regression models and life tables (with discussion)}.
\bjournal{Journal of the Royal Statistical Society}
\bvolume{34}
\bpages{187--220}.
\end{barticle}
\endbibitem

\bibitem[\protect\citeauthoryear{Feinberg, Ohlsson and
  Henikoff}{2006}]{feinberg2006epigenetic}
\begin{barticle}[author]
\bauthor{\bsnm{Feinberg},~\bfnm{A.~P.}\binits{A.~P.}},
  \bauthor{\bsnm{Ohlsson},~\bfnm{R.}\binits{R.}} \AND
  \bauthor{\bsnm{Henikoff},~\bfnm{S.}\binits{S.}}
(\byear{2006}).
\btitle{The epigenetic progenitor origin of human cancer}.
\bjournal{Nature Reviews Genetics}
\bvolume{7}
\bpages{21-33}.
\end{barticle}
\endbibitem

\bibitem[\protect\citeauthoryear{Fleischer et~al.}{2014}]{fleischer2014genome}
\begin{barticle}[author]
\bauthor{\bsnm{Fleischer},~\bfnm{Thomas}\binits{T.}},
  \bauthor{\bsnm{Frigessi},~\bfnm{Arnoldo}\binits{A.}},
  \bauthor{\bsnm{Johnson},~\bfnm{Kevin~C}\binits{K.~C.}},
  \bauthor{\bsnm{Edvardsen},~\bfnm{Hege}\binits{H.}},
  \bauthor{\bsnm{Touleimat},~\bfnm{Nizar}\binits{N.}},
  \bauthor{\bsnm{Klajic},~\bfnm{Jovana}\binits{J.}},
  \bauthor{\bsnm{Riis},~\bfnm{Margit~LH}\binits{M.~L.}},
  \bauthor{\bsnm{Haakensen},~\bfnm{Vilde}\binits{V.}},
  \bauthor{\bsnm{W{\"a}rnberg},~\bfnm{Fredrik}\binits{F.}},
  \bauthor{\bsnm{Naume},~\bfnm{Bj{\o}rn}\binits{B.}} \betal{et~al.}
(\byear{2014}).
\btitle{Genome-wide {DNA} methylation profiles in progression to in situ and
  invasive carcinoma of the breast with impact on gene transcription and
  prognosis}.
\bjournal{Genome Biol}
\bvolume{15}
\bpages{435}.
\end{barticle}
\endbibitem

\bibitem[\protect\citeauthoryear{Gao et~al.}{2013}]{gao2013dna}
\begin{barticle}[author]
\bauthor{\bsnm{Gao},~\bfnm{Fan}\binits{F.}},
  \bauthor{\bsnm{Shi},~\bfnm{Lingling}\binits{L.}},
  \bauthor{\bsnm{Russin},~\bfnm{Jonathan}\binits{J.}},
  \bauthor{\bsnm{Zeng},~\bfnm{Liyun}\binits{L.}},
  \bauthor{\bsnm{Chang},~\bfnm{Xiao}\binits{X.}},
  \bauthor{\bsnm{He},~\bfnm{Shuhan}\binits{S.}},
  \bauthor{\bsnm{Chen},~\bfnm{Thomas~C}\binits{T.~C.}},
  \bauthor{\bsnm{Giannotta},~\bfnm{Steven~L}\binits{S.~L.}},
  \bauthor{\bsnm{Weisenberger},~\bfnm{Daniel~J}\binits{D.~J.}},
  \bauthor{\bsnm{Zada},~\bfnm{Gabriel}\binits{G.}} \betal{et~al.}
(\byear{2013}).
\btitle{{DNA} methylation in the malignant transformation of meningiomas}.
\bjournal{PloS One}
\bvolume{8}
\bpages{e54114}.
\end{barticle}
\endbibitem

\bibitem[\protect\citeauthoryear{Girvan and Newman}{2002}]{girvan2002community}
\begin{barticle}[author]
\bauthor{\bsnm{Girvan},~\bfnm{Michelle}\binits{M.}} \AND
  \bauthor{\bsnm{Newman},~\bfnm{Mark~EJ}\binits{M.~E.}}
(\byear{2002}).
\btitle{Community structure in social and biological networks}.
\bjournal{Proceedings of the National Academy of Sciences}
\bvolume{99}
\bpages{7821--7826}.
\end{barticle}
\endbibitem

\bibitem[\protect\citeauthoryear{Hampton}{2006}]{hampton2006cancer}
\begin{barticle}[author]
\bauthor{\bsnm{Hampton},~\bfnm{Tracy}\binits{T.}}
(\byear{2006}).
\btitle{Cancer genome atlas}.
\bjournal{JAMA: The Journal of the American Medical Association}
\bvolume{296}
\bpages{1958--1958}.
\end{barticle}
\endbibitem

\bibitem[\protect\citeauthoryear{Harrell}{2001}]{harrell2001regression}
\begin{bbook}[author]
\bauthor{\bsnm{Harrell},~\bfnm{Frank~E}\binits{F.~E.}}
(\byear{2001}).
\btitle{Regression modeling strategies: with applications to linear models,
  logistic regression, and survival analysis}.
\bpublisher{Springer}.
\end{bbook}
\endbibitem

\bibitem[\protect\citeauthoryear{Holland, Laskey and
  Leinhardt}{1983}]{holland1983stochastic}
\begin{barticle}[author]
\bauthor{\bsnm{Holland},~\bfnm{Paul~W}\binits{P.~W.}},
  \bauthor{\bsnm{Laskey},~\bfnm{Kathryn~Blackmond}\binits{K.~B.}} \AND
  \bauthor{\bsnm{Leinhardt},~\bfnm{Samuel}\binits{S.}}
(\byear{1983}).
\btitle{Stochastic blockmodels: First steps}.
\bjournal{Social Networks}
\bvolume{5}
\bpages{109--137}.
\end{barticle}
\endbibitem

\bibitem[\protect\citeauthoryear{Hotelling}{1936}]{hotelling1936relations}
\begin{barticle}[author]
\bauthor{\bsnm{Hotelling},~\bfnm{Harold}\binits{H.}}
(\byear{1936}).
\btitle{Relations between two sets of variates}.
\bjournal{Biometrika}
\bvolume{28}
\bpages{321--377}.
\end{barticle}
\endbibitem

\bibitem[\protect\citeauthoryear{Jacob et~al.}{2012}]{jacob2012more}
\begin{barticle}[author]
\bauthor{\bsnm{Jacob},~\bfnm{Laurent}\binits{L.}},
  \bauthor{\bsnm{Neuvial},~\bfnm{Pierre}\binits{P.}},
  \bauthor{\bsnm{Dudoit},~\bfnm{Sandrine}\binits{S.}} \betal{et~al.}
(\byear{2012}).
\btitle{More power via graph-structured tests for differential expression of
  gene networks}.
\bjournal{The Annals of Applied Statistics}
\bvolume{6}
\bpages{561--600}.
\end{barticle}
\endbibitem

\bibitem[\protect\citeauthoryear{Johnstone and
  Silverman}{2004}]{johnstone2004needles}
\begin{barticle}[author]
\bauthor{\bsnm{Johnstone},~\bfnm{Iain~M}\binits{I.~M.}} \AND
  \bauthor{\bsnm{Silverman},~\bfnm{Bernard~W}\binits{B.~W.}}
(\byear{2004}).
\btitle{Needles and straw in haystacks: Empirical Bayes estimates of possibly
  sparse sequences}.
\bjournal{Annals of Statistics}
\bpages{1594--1649}.
\end{barticle}
\endbibitem

\bibitem[\protect\citeauthoryear{Jones}{2012}]{jones2012functions}
\begin{barticle}[author]
\bauthor{\bsnm{Jones},~\bfnm{P.~A.}\binits{P.~A.}}
(\byear{2012}).
\btitle{Functions of {DNA} methylation: islands, start sites, gene bodies and
  beyond}.
\bjournal{Nature Reviews Genetics}
\bvolume{13}
\bpages{484--492}.
\end{barticle}
\endbibitem

\bibitem[\protect\citeauthoryear{Kang et~al.}{2001}]{kang2001cpg}
\begin{barticle}[author]
\bauthor{\bsnm{Kang},~\bfnm{Gyeong~Hoon}\binits{G.~H.}},
  \bauthor{\bsnm{Shim},~\bfnm{Yhong-Hee}\binits{Y.-H.}},
  \bauthor{\bsnm{Jung},~\bfnm{Hwoon-Yong}\binits{H.-Y.}},
  \bauthor{\bsnm{Kim},~\bfnm{Woo~Ho}\binits{W.~H.}},
  \bauthor{\bsnm{Ro},~\bfnm{Jae~Y}\binits{J.~Y.}} \AND
  \bauthor{\bsnm{Rhyu},~\bfnm{Mun-Gan}\binits{M.-G.}}
(\byear{2001}).
\btitle{{CpG} island methylation in premalignant stages of gastric carcinoma}.
\bjournal{Cancer Research}
\bvolume{61}
\bpages{2847--2851}.
\end{barticle}
\endbibitem

\bibitem[\protect\citeauthoryear{Kang et~al.}{2003}]{kang2003profile}
\begin{barticle}[author]
\bauthor{\bsnm{Kang},~\bfnm{Gyeong~Hoon}\binits{G.~H.}},
  \bauthor{\bsnm{Lee},~\bfnm{Sun}\binits{S.}},
  \bauthor{\bsnm{Kim},~\bfnm{Jung-Sun}\binits{J.-S.}} \AND
  \bauthor{\bsnm{Jung},~\bfnm{Hwoon-Yong}\binits{H.-Y.}}
(\byear{2003}).
\btitle{Profile of aberrant {CpG} island methylation along multistep gastric
  carcinogenesis}.
\bjournal{Laboratory Investigation}
\bvolume{83}
\bpages{519--526}.
\end{barticle}
\endbibitem

\bibitem[\protect\citeauthoryear{Katenka et~al.}{2012}]{katenka2012inference}
\begin{barticle}[author]
\bauthor{\bsnm{Katenka},~\bfnm{Natallia}\binits{N.}},
  \bauthor{\bsnm{Kolaczyk},~\bfnm{Eric~D}\binits{E.~D.}} \betal{et~al.}
(\byear{2012}).
\btitle{Inference and characterization of multi-attribute networks with
  application to computational biology}.
\bjournal{The Annals of Applied Statistics}
\bvolume{6}
\bpages{1068--1094}.
\end{barticle}
\endbibitem

\bibitem[\protect\citeauthoryear{Kishida et~al.}{2012}]{kishida2012epigenetic}
\begin{barticle}[author]
\bauthor{\bsnm{Kishida},~\bfnm{Yugo}\binits{Y.}},
  \bauthor{\bsnm{Natsume},~\bfnm{Atsushi}\binits{A.}},
  \bauthor{\bsnm{Kondo},~\bfnm{Yutaka}\binits{Y.}},
  \bauthor{\bsnm{Takeuchi},~\bfnm{Ichiro}\binits{I.}},
  \bauthor{\bsnm{An},~\bfnm{Byonggu}\binits{B.}},
  \bauthor{\bsnm{Okamoto},~\bfnm{Yasuyuki}\binits{Y.}},
  \bauthor{\bsnm{Shinjo},~\bfnm{Keiko}\binits{K.}},
  \bauthor{\bsnm{Saito},~\bfnm{Kiyoshi}\binits{K.}},
  \bauthor{\bsnm{Ando},~\bfnm{Hitoshi}\binits{H.}},
  \bauthor{\bsnm{Ohka},~\bfnm{Fumiharu}\binits{F.}} \betal{et~al.}
(\byear{2012}).
\btitle{Epigenetic subclassification of meningiomas based on genome-wide {DNA}
  methylation analyses}.
\bjournal{Carcinogenesis}
\bvolume{33}
\bpages{436--441}.
\end{barticle}
\endbibitem

\bibitem[\protect\citeauthoryear{Lai and Shiekhattar}{2014}]{lai2014long}
\begin{barticle}[author]
\bauthor{\bsnm{Lai},~\bfnm{Fan}\binits{F.}} \AND
  \bauthor{\bsnm{Shiekhattar},~\bfnm{Ramin}\binits{R.}}
(\byear{2014}).
\btitle{Where long noncoding RNAs meet DNA methylation.}
\bjournal{Cell Research}
\bvolume{24}
\bpages{263--264}.
\end{barticle}
\endbibitem

\bibitem[\protect\citeauthoryear{Latouche
  et~al.}{2011}]{latouche2011overlapping}
\begin{barticle}[author]
\bauthor{\bsnm{Latouche},~\bfnm{Pierre}\binits{P.}},
  \bauthor{\bsnm{Birmel{\'e}},~\bfnm{Etienne}\binits{E.}},
  \bauthor{\bsnm{Ambroise},~\bfnm{Christophe}\binits{C.}} \betal{et~al.}
(\byear{2011}).
\btitle{Overlapping stochastic block models with application to the french
  political blogosphere}.
\bjournal{The Annals of Applied Statistics}
\bvolume{5}
\bpages{309--336}.
\end{barticle}
\endbibitem

\bibitem[\protect\citeauthoryear{Li and Li}{2010}]{li2010variable}
\begin{barticle}[author]
\bauthor{\bsnm{Li},~\bfnm{Caiyan}\binits{C.}} \AND
  \bauthor{\bsnm{Li},~\bfnm{Hongzhe}\binits{H.}}
(\byear{2010}).
\btitle{Variable selection and regression analysis for graph-structured
  covariates with an application to genomics}.
\bjournal{The Annals of Applied Statistics}
\bvolume{4}
\bpages{1498}.
\end{barticle}
\endbibitem

\bibitem[\protect\citeauthoryear{Li and Wang}{2014}]{li2014quantifying}
\begin{barticle}[author]
\bauthor{\bsnm{Li},~\bfnm{Chunhe}\binits{C.}} \AND
  \bauthor{\bsnm{Wang},~\bfnm{Jin}\binits{J.}}
(\byear{2014}).
\btitle{Quantifying the underlying landscape and paths of cancer}.
\bjournal{Journal of The Royal Society Interface}
\bvolume{11}
\bpages{20140774}.
\end{barticle}
\endbibitem

\bibitem[\protect\citeauthoryear{Luo et~al.}{2014}]{luo2014differences}
\begin{barticle}[author]
\bauthor{\bsnm{Luo},~\bfnm{Yanxin}\binits{Y.}},
  \bauthor{\bsnm{Wong},~\bfnm{Chao-Jen}\binits{C.-J.}},
  \bauthor{\bsnm{Kaz},~\bfnm{Andrew~M}\binits{A.~M.}},
  \bauthor{\bsnm{Dzieciatkowski},~\bfnm{Slavomir}\binits{S.}},
  \bauthor{\bsnm{Carter},~\bfnm{Kelly~T}\binits{K.~T.}},
  \bauthor{\bsnm{Morris},~\bfnm{Shelli~M}\binits{S.~M.}},
  \bauthor{\bsnm{Wang},~\bfnm{Jianping}\binits{J.}},
  \bauthor{\bsnm{Willis},~\bfnm{Joseph~E}\binits{J.~E.}},
  \bauthor{\bsnm{Makar},~\bfnm{Karen~W}\binits{K.~W.}},
  \bauthor{\bsnm{Ulrich},~\bfnm{Cornelia~M}\binits{C.~M.}} \betal{et~al.}
(\byear{2014}).
\btitle{Differences in DNA methylation signatures reveal multiple pathways of
  progression from adenoma to colorectal cancer}.
\bjournal{Gastroenterology}
\bvolume{147}
\bpages{418--429}.
\end{barticle}
\endbibitem

\bibitem[\protect\citeauthoryear{Maekawa et~al.}{2013}]{maekawa2013genome}
\begin{barticle}[author]
\bauthor{\bsnm{Maekawa},~\bfnm{Ryo}\binits{R.}},
  \bauthor{\bsnm{Sato},~\bfnm{Shun}\binits{S.}},
  \bauthor{\bsnm{Yamagata},~\bfnm{Yoshiaki}\binits{Y.}},
  \bauthor{\bsnm{Asada},~\bfnm{Hiromi}\binits{H.}},
  \bauthor{\bsnm{Tamura},~\bfnm{Isao}\binits{I.}},
  \bauthor{\bsnm{Lee},~\bfnm{Lifa}\binits{L.}},
  \bauthor{\bsnm{Okada},~\bfnm{Maki}\binits{M.}},
  \bauthor{\bsnm{Tamura},~\bfnm{Hiroshi}\binits{H.}},
  \bauthor{\bsnm{Takaki},~\bfnm{Eiichi}\binits{E.}},
  \bauthor{\bsnm{Nakai},~\bfnm{Akira}\binits{A.}} \betal{et~al.}
(\byear{2013}).
\btitle{Genome-wide {DNA} methylation analysis reveals a potential mechanism
  for the pathogenesis and development of uterine leiomyomas}.
\bjournal{PloS One}
\bvolume{8}
\bpages{e66632}.
\end{barticle}
\endbibitem

\bibitem[\protect\citeauthoryear{Mardia}{2013}]{mardia2013statistical}
\begin{barticle}[author]
\bauthor{\bsnm{Mardia},~\bfnm{Kanti~V}\binits{K.~V.}}
(\byear{2013}).
\btitle{Statistical approaches to three key challenges in protein structural
  bioinformatics}.
\bjournal{Journal of the Royal Statistical Society: Series C (Applied
  Statistics)}
\bvolume{62}
\bpages{487--514}.
\end{barticle}
\endbibitem

\bibitem[\protect\citeauthoryear{Nandi, Sumana and
  Bhattacharya}{2014}]{nandi2014social}
\begin{barticle}[author]
\bauthor{\bsnm{Nandi},~\bfnm{Anjan~K}\binits{A.~K.}},
  \bauthor{\bsnm{Sumana},~\bfnm{Annagiri}\binits{A.}} \AND
  \bauthor{\bsnm{Bhattacharya},~\bfnm{Kunal}\binits{K.}}
(\byear{2014}).
\btitle{Social insect colony as a biological regulatory system: modelling
  information flow in dominance networks}.
\bjournal{Journal of The Royal Society Interface}
\bvolume{11}
\bpages{20140951}.
\end{barticle}
\endbibitem

\bibitem[\protect\citeauthoryear{Navarro et~al.}{2012}]{navarro2012genome}
\begin{barticle}[author]
\bauthor{\bsnm{Navarro},~\bfnm{Antonia}\binits{A.}},
  \bauthor{\bsnm{Yin},~\bfnm{Ping}\binits{P.}},
  \bauthor{\bsnm{Monsivais},~\bfnm{Diana}\binits{D.}},
  \bauthor{\bsnm{Lin},~\bfnm{Simon~M}\binits{S.~M.}},
  \bauthor{\bsnm{Du},~\bfnm{Pan}\binits{P.}},
  \bauthor{\bsnm{Wei},~\bfnm{Jian-Jun}\binits{J.-J.}} \AND
  \bauthor{\bsnm{Bulun},~\bfnm{Serdar~E}\binits{S.~E.}}
(\byear{2012}).
\btitle{Genome-wide {DNA} methylation indicates silencing of tumor suppressor
  genes in uterine leiomyoma}.
\bjournal{PloS One}
\bvolume{7}
\bpages{e33284}.
\end{barticle}
\endbibitem

\bibitem[\protect\citeauthoryear{Newman}{2004}]{newman2004detecting}
\begin{barticle}[author]
\bauthor{\bsnm{Newman},~\bfnm{Mark~EJ}\binits{M.~E.}}
(\byear{2004}).
\btitle{Detecting community structure in networks}.
\bjournal{The European Physical Journal B-Condensed Matter and Complex Systems}
\bvolume{38}
\bpages{321--330}.
\end{barticle}
\endbibitem

\bibitem[\protect\citeauthoryear{Newman and Girvan}{2004}]{newman2004finding}
\begin{barticle}[author]
\bauthor{\bsnm{Newman},~\bfnm{Mark~EJ}\binits{M.~E.}} \AND
  \bauthor{\bsnm{Girvan},~\bfnm{Michelle}\binits{M.}}
(\byear{2004}).
\btitle{Finding and evaluating community structure in networks}.
\bjournal{Physical Review E}
\bvolume{69}
\bpages{026113}.
\end{barticle}
\endbibitem

\bibitem[\protect\citeauthoryear{Olhede and Wolfe}{2014}]{olhede2014net}
\begin{barticle}[author]
\bauthor{\bsnm{Olhede},~\bfnm{Sofia~C.}\binits{S.~C.}} \AND
  \bauthor{\bsnm{Wolfe},~\bfnm{Patrick~J.}\binits{P.~J.}}
(\byear{2014}).
\btitle{Network histograms and universality of blockmodel approximation}.
\bjournal{Proceedings of the National Academy of Sciences}
\bvolume{111}
\bpages{14722-14727}.
\bdoi{10.1073/pnas.1400374111}
\end{barticle}
\endbibitem

\bibitem[\protect\citeauthoryear{Palla, Lov{\'a}sz and
  Vicsek}{2010}]{palla2010multifractal}
\begin{barticle}[author]
\bauthor{\bsnm{Palla},~\bfnm{Gergely}\binits{G.}},
  \bauthor{\bsnm{Lov{\'a}sz},~\bfnm{L{\'a}szl{\'o}}\binits{L.}} \AND
  \bauthor{\bsnm{Vicsek},~\bfnm{Tam{\'a}s}\binits{T.}}
(\byear{2010}).
\btitle{Multifractal network generator}.
\bjournal{Proceedings of the National Academy of Sciences}
\bvolume{107}
\bpages{7640--7645}.
\end{barticle}
\endbibitem

\bibitem[\protect\citeauthoryear{Peng et~al.}{2010}]{peng2010regularized}
\begin{barticle}[author]
\bauthor{\bsnm{Peng},~\bfnm{Jie}\binits{J.}},
  \bauthor{\bsnm{Zhu},~\bfnm{Ji}\binits{J.}},
  \bauthor{\bsnm{Bergamaschi},~\bfnm{Anna}\binits{A.}},
  \bauthor{\bsnm{Han},~\bfnm{Wonshik}\binits{W.}},
  \bauthor{\bsnm{Noh},~\bfnm{Dong-Young}\binits{D.-Y.}},
  \bauthor{\bsnm{Pollack},~\bfnm{Jonathan~R}\binits{J.~R.}} \AND
  \bauthor{\bsnm{Wang},~\bfnm{Pei}\binits{P.}}
(\byear{2010}).
\btitle{Regularized multivariate regression for identifying master predictors
  with application to integrative genomics study of breast cancer}.
\bjournal{The Annals of Applied Statistics}
\bvolume{4}
\bpages{53}.
\end{barticle}
\endbibitem

\bibitem[\protect\citeauthoryear{Qin and Rohe}{2013}]{qin2013regularized}
\begin{binproceedings}[author]
\bauthor{\bsnm{Qin},~\bfnm{Tai}\binits{T.}} \AND
  \bauthor{\bsnm{Rohe},~\bfnm{Karl}\binits{K.}}
(\byear{2013}).
\btitle{Regularized spectral clustering under the degree-corrected stochastic
  blockmodel}.
In \bbooktitle{Advances in Neural Information Processing Systems}
\bpages{3120--3128}.
\end{binproceedings}
\endbibitem

\bibitem[\protect\citeauthoryear{Reznik, Watson and
  Chaudhary}{2013}]{reznik2013stubborn}
\begin{barticle}[author]
\bauthor{\bsnm{Reznik},~\bfnm{Ed}\binits{E.}},
  \bauthor{\bsnm{Watson},~\bfnm{Alex}\binits{A.}} \AND
  \bauthor{\bsnm{Chaudhary},~\bfnm{Osman}\binits{O.}}
(\byear{2013}).
\btitle{The stubborn roots of metabolic cycles}.
\bjournal{Journal of The Royal Society Interface}
\bvolume{10}
\bpages{20130087}.
\end{barticle}
\endbibitem

\bibitem[\protect\citeauthoryear{Riolo and Newman}{2012}]{riolo2012first}
\begin{barticle}[author]
\bauthor{\bsnm{Riolo},~\bfnm{Maria~A}\binits{M.~A.}} \AND
  \bauthor{\bsnm{Newman},~\bfnm{MEJ}\binits{M.}}
(\byear{2012}).
\btitle{First-principles multiway spectral partitioning of graphs}.
\bjournal{arXiv preprint arXiv:1209.5969}.
\end{barticle}
\endbibitem

\bibitem[\protect\citeauthoryear{Saavedra
  et~al.}{2014}]{saavedra2014structurally}
\begin{barticle}[author]
\bauthor{\bsnm{Saavedra},~\bfnm{Serguei}\binits{S.}},
  \bauthor{\bsnm{Rohr},~\bfnm{Rudolf~P}\binits{R.~P.}},
  \bauthor{\bsnm{Gilarranz},~\bfnm{Luis~J}\binits{L.~J.}} \AND
  \bauthor{\bsnm{Bascompte},~\bfnm{Jordi}\binits{J.}}
(\byear{2014}).
\btitle{How structurally stable are global socioeconomic systems?}
\bjournal{Journal of The Royal Society Interface}
\bvolume{11}
\bpages{20140693}.
\end{barticle}
\endbibitem

\bibitem[\protect\citeauthoryear{Shen-Orr et~al.}{2002}]{shen2002network}
\begin{barticle}[author]
\bauthor{\bsnm{Shen-Orr},~\bfnm{Shai~S}\binits{S.~S.}},
  \bauthor{\bsnm{Milo},~\bfnm{Ron}\binits{R.}},
  \bauthor{\bsnm{Mangan},~\bfnm{Shmoolik}\binits{S.}} \AND
  \bauthor{\bsnm{Alon},~\bfnm{Uri}\binits{U.}}
(\byear{2002}).
\btitle{Network motifs in the transcriptional regulation network of Escherichia
  coli}.
\bjournal{Nature Genetics}
\bvolume{31}
\bpages{64--68}.
\end{barticle}
\endbibitem

\bibitem[\protect\citeauthoryear{Taylor et~al.}{2009}]{taylor2009dynamic}
\begin{barticle}[author]
\bauthor{\bsnm{Taylor},~\bfnm{Ian~W}\binits{I.~W.}},
  \bauthor{\bsnm{Linding},~\bfnm{Rune}\binits{R.}},
  \bauthor{\bsnm{Warde-Farley},~\bfnm{David}\binits{D.}},
  \bauthor{\bsnm{Liu},~\bfnm{Yongmei}\binits{Y.}},
  \bauthor{\bsnm{Pesquita},~\bfnm{Catia}\binits{C.}},
  \bauthor{\bsnm{Faria},~\bfnm{Daniel}\binits{D.}},
  \bauthor{\bsnm{Bull},~\bfnm{Shelley}\binits{S.}},
  \bauthor{\bsnm{Pawson},~\bfnm{Tony}\binits{T.}},
  \bauthor{\bsnm{Morris},~\bfnm{Quaid}\binits{Q.}} \AND
  \bauthor{\bsnm{Wrana},~\bfnm{Jeffrey~L}\binits{J.~L.}}
(\byear{2009}).
\btitle{Dynamic modularity in protein interaction networks predicts breast
  cancer outcome}.
\bjournal{Nature Biotechnology}
\bvolume{27}
\bpages{199--204}.
\end{barticle}
\endbibitem

\bibitem[\protect\citeauthoryear{Tran and Kwon}{2013}]{tran2013relationship}
\begin{barticle}[author]
\bauthor{\bsnm{Tran},~\bfnm{Tien-Dzung}\binits{T.-D.}} \AND
  \bauthor{\bsnm{Kwon},~\bfnm{Yung-Keun}\binits{Y.-K.}}
(\byear{2013}).
\btitle{The relationship between modularity and robustness in signalling
  networks}.
\bjournal{Journal of The Royal Society Interface}
\bvolume{10}
\bpages{20130771}.
\end{barticle}
\endbibitem

\bibitem[\protect\citeauthoryear{Van~Hoesel et~al.}{2013}]{van2013assessment}
\begin{barticle}[author]
\bauthor{\bsnm{Van~Hoesel},~\bfnm{AQ}\binits{A.}},
  \bauthor{\bsnm{Sato},~\bfnm{Y}\binits{Y.}},
  \bauthor{\bsnm{Elashoff},~\bfnm{DA}\binits{D.}},
  \bauthor{\bsnm{Turner},~\bfnm{RR}\binits{R.}},
  \bauthor{\bsnm{Giuliano},~\bfnm{AE}\binits{A.}},
  \bauthor{\bsnm{Shamonki},~\bfnm{JM}\binits{J.}},
  \bauthor{\bsnm{Kuppen},~\bfnm{PJK}\binits{P.}}, \bauthor{\bparticle{van~de}
  \bsnm{Velde},~\bfnm{CJH}\binits{C.}} \AND
  \bauthor{\bsnm{Hoon},~\bfnm{DSB}\binits{D.}}
(\byear{2013}).
\btitle{Assessment of {DNA} methylation status in early stages of breast cancer
  development}.
\bjournal{British Journal of Cancer}
\bvolume{108}
\bpages{2033--2038}.
\end{barticle}
\endbibitem

\bibitem[\protect\citeauthoryear{Venters and Pugh}{2013}]{venters2013genomic}
\begin{barticle}[author]
\bauthor{\bsnm{Venters},~\bfnm{Bryan~J}\binits{B.~J.}} \AND
  \bauthor{\bsnm{Pugh},~\bfnm{B~Franklin}\binits{B.~F.}}
(\byear{2013}).
\btitle{Genomic organization of human transcription initiation complexes}.
\bjournal{Nature}
\bvolume{502}
\bpages{53--58}.
\end{barticle}
\endbibitem

\bibitem[\protect\citeauthoryear{Verschuur-Maes, de~Bruin and van
  Diest}{2012}]{verschuur2012epigenetic}
\begin{barticle}[author]
\bauthor{\bsnm{Verschuur-Maes},~\bfnm{Anoek~HJ}\binits{A.~H.}},
  \bauthor{\bparticle{de} \bsnm{Bruin},~\bfnm{Peter~C}\binits{P.~C.}} \AND
  \bauthor{\bparticle{van} \bsnm{Diest},~\bfnm{Paul~J}\binits{P.~J.}}
(\byear{2012}).
\btitle{Epigenetic progression of columnar cell lesions of the breast to
  invasive breast cancer}.
\bjournal{Breast Cancer Research and Treatment}
\bvolume{136}
\bpages{705--715}.
\end{barticle}
\endbibitem

\bibitem[\protect\citeauthoryear{Vu et~al.}{2013}]{vu2013model}
\begin{barticle}[author]
\bauthor{\bsnm{Vu},~\bfnm{Duy~Q}\binits{D.~Q.}},
  \bauthor{\bsnm{Hunter},~\bfnm{David~R}\binits{D.~R.}},
  \bauthor{\bsnm{Schweinberger},~\bfnm{Michael}\binits{M.}} \betal{et~al.}
(\byear{2013}).
\btitle{Model-based clustering of large networks}.
\bjournal{The Annals of Applied Statistics}
\bvolume{7}
\bpages{1010--1039}.
\end{barticle}
\endbibitem

\bibitem[\protect\citeauthoryear{Wagner}{2002}]{wagner2002estimating}
\begin{barticle}[author]
\bauthor{\bsnm{Wagner},~\bfnm{Andreas}\binits{A.}}
(\byear{2002}).
\btitle{Estimating coarse gene network structure from large-scale gene
  perturbation data}.
\bjournal{Genome Research}
\bvolume{12}
\bpages{309--315}.
\end{barticle}
\endbibitem

\bibitem[\protect\citeauthoryear{Wei and Pan}{2010}]{wei2010network}
\begin{barticle}[author]
\bauthor{\bsnm{Wei},~\bfnm{Peng}\binits{P.}} \AND
  \bauthor{\bsnm{Pan},~\bfnm{Wei}\binits{W.}}
(\byear{2010}).
\btitle{Network-based genomic discovery: application and comparison of Markov
  random-field models}.
\bjournal{Journal of the Royal Statistical Society: Series C (Applied
  Statistics)}
\bvolume{59}
\bpages{105--125}.
\end{barticle}
\endbibitem

\bibitem[\protect\citeauthoryear{Xie et~al.}{2013}]{xie2013epigenomic}
\begin{barticle}[author]
\bauthor{\bsnm{Xie},~\bfnm{Wei}\binits{W.}},
  \bauthor{\bsnm{Schultz},~\bfnm{Matthew~D}\binits{M.~D.}},
  \bauthor{\bsnm{Lister},~\bfnm{Ryan}\binits{R.}},
  \bauthor{\bsnm{Hou},~\bfnm{Zhonggang}\binits{Z.}},
  \bauthor{\bsnm{Rajagopal},~\bfnm{Nisha}\binits{N.}},
  \bauthor{\bsnm{Ray},~\bfnm{Pradipta}\binits{P.}},
  \bauthor{\bsnm{Whitaker},~\bfnm{John~W}\binits{J.~W.}},
  \bauthor{\bsnm{Tian},~\bfnm{Shulan}\binits{S.}},
  \bauthor{\bsnm{Hawkins},~\bfnm{R~David}\binits{R.~D.}},
  \bauthor{\bsnm{Leung},~\bfnm{Danny}\binits{D.}} \betal{et~al.}
(\byear{2013}).
\btitle{Epigenomic analysis of multilineage differentiation of human embryonic
  stem cells}.
\bjournal{Cell}
\bvolume{153}
\bpages{1134--1148}.
\end{barticle}
\endbibitem

\bibitem[\protect\citeauthoryear{Yamamoto et~al.}{2012}]{yamamoto2012molecular}
\begin{barticle}[author]
\bauthor{\bsnm{Yamamoto},~\bfnm{Eiichiro}\binits{E.}},
  \bauthor{\bsnm{Suzuki},~\bfnm{Hiromu}\binits{H.}},
  \bauthor{\bsnm{Yamano},~\bfnm{Hiro-o}\binits{H.-o.}},
  \bauthor{\bsnm{Maruyama},~\bfnm{Reo}\binits{R.}},
  \bauthor{\bsnm{Nojima},~\bfnm{Masanori}\binits{M.}},
  \bauthor{\bsnm{Kamimae},~\bfnm{Seiko}\binits{S.}},
  \bauthor{\bsnm{Sawada},~\bfnm{Takeshi}\binits{T.}},
  \bauthor{\bsnm{Ashida},~\bfnm{Masami}\binits{M.}},
  \bauthor{\bsnm{Yoshikawa},~\bfnm{Kenjiro}\binits{K.}},
  \bauthor{\bsnm{Kimura},~\bfnm{Tomoaki}\binits{T.}} \betal{et~al.}
(\byear{2012}).
\btitle{Molecular dissection of premalignant colorectal lesions reveals early
  onset of the {CpG} island methylator phenotype}.
\bjournal{The American Journal of Pathology}
\bvolume{181}
\bpages{1847--1861}.
\end{barticle}
\endbibitem

\end{thebibliography}

\section*{Acknowledgements}
We are very grateful to Professor S.C. Olhede for literature suggestions, fruitful discussions and helpful comments.

% AOS,AOAS: If there are supplements please fill:
\begin{supplement}[id=supp]
  \sname{Supplement}
  \stitle{Supplementary Tables and Figures}
  \slink[doi]{}
   \sdatatype{.pdf}
  \sdescription{Supplementary Tables S1-S5 and Supplementary Figures S1-S2}
\end{supplement}

\end{document}